\DeclareUnicodeCharacter{03B1}{\ensuremath{\alpha}}
\DeclareUnicodeCharacter{03B2}{\ensuremath{\alpha}}
\PassOptionsToPackage{unicode}{hyperref}
\PassOptionsToPackage{hyphens}{url}
\documentclass[]{elsarticle}
\usepackage{amsmath,amssymb}
\usepackage{iftex}
\ifPDFTeX
\usepackage[T1]{fontenc}
\usepackage[utf8]{inputenc}
\usepackage{textcomp}
\else
\usepackage{unicode-math}
\defaultfontfeatures{Scale=MatchLowercase}\defaultfontfeatures[\rmfamily]{Ligatures=TeX,Scale=1}\fi\usepackage{lmodern}\ifPDFTeX\else\fi\IfFileExists{upquote.sty}{\usepackage{upquote}}{}\IfFileExists{microtype.sty}{\usepackage[]{microtype}\UseMicrotypeSet[protrusion]{basicmath}}{}\usepackage{xcolor}\usepackage[left=2.5cm,right=2.5cm,top=2.5cm,bottom=2.5cm]{geometry}\usepackage{longtable,booktabs,array}\usepackage{calc}\usepackage{etoolbox}\makeatletter\patchcmd\longtable{\par}{\if@noskipsec\mbox{}\fi\par}{}{}\makeatother\IfFileExists{footnotehyper.sty}{\usepackage{footnotehyper}}{\usepackage{footnote}}\makesavenoteenv{longtable}\usepackage{graphicx}\makeatletter\def\maxwidth{\ifdim\Gin@nat@width>\linewidth\linewidth\else\Gin@nat@width\fi}\def\maxheight{\ifdim\Gin@nat@height>\textheight\textheight\else\Gin@nat@height\fi}\makeatother\setkeys{Gin}{width=\maxwidth,height=\maxheight,keepaspectratio}\makeatletter\def\fps@figure{htbp}\makeatother\providecommand{\tightlist}{\setlength{\itemsep}{0pt}\setlength{\parskip}{0pt}}\newlength{\cslhangindent}\newlength{\csllabelwidth}\newlength{\cslentryspacingunit}\usepackage{calc}\usepackage[fontsize=11pt]{scrextend}\usepackage{setspace}\usepackage{placeins}\usepackage{soul}\usepackage{color}\usepackage{floatrow}\usepackage{float}\usepackage{fancyhdr}\usepackage{booktabs}\usepackage{longtable}\usepackage{array}\usepackage{multirow}\usepackage{wrapfig}\usepackage{float}\usepackage{colortbl}\usepackage{pdflscape}\usepackage{tabu}\usepackage{threeparttable}\usepackage{threeparttablex}\usepackage[normalem]{ulem}\usepackage{makecell}\usepackage{xcolor}\ifLuaTeX\usepackage{selnolig}\fi\IfFileExists{bookmark.sty}{\usepackage{bookmark}}{\usepackage{hyperref}}\IfFileExists{xurl.sty}{\usepackage{xurl}}{}\hypersetup{hidelinks,pdfcreator={LaTeX via pandoc}}\author{}\date{\vspace{-2.5em}}

\makeatletter
\def\ps@pprintTitle{%
     \let\@oddhead\@empty
     \let\@evenhead\@empty
     \def\@oddfoot{\footnotesize\itshape
       Preprint submitted to ...\hfill\today}%
     \let\@evenfoot\@oddfoot}
\makeatother

\begin{document}

\begin{frontmatter}

\title{LSTM-ARIMA as a Hybrid Approach in Algorithmic Investment Strategies}

\vspace{1\baselineskip} 

\author[WNEUW]{Kamil Kashif\fnref{1}}
   
\author[WNEUW2]{Robert Ślepaczuk\corref{cor1}\fnref{2}}
\ead{rslepaczuk@wne.uw.edu.pl} 
   
\affiliation[WNEUW]{Quantitative Finance Research Group, University of Warsaw, Faculty of Economic Sciences, ul. Dluga 44-50, 00-241, Warsaw, Poland}

\affiliation[WNEUW2]{Quantitative Finance Research Group, Department of Quantitative Finance, Faculty of Economic Sciences, University of Warsaw, ul. Dluga 44-50, 00-241, Warsaw, Poland}

\cortext[cor1]{Corresponding author}

\fntext[1]{ORCID: https://orcid.org/0009-0004-0666-581X}
\fntext[2]{ORCID: https://orcid.org/0000-0001-5227-2014}
  
  \begin{abstract}
  This study focuses on building an algorithmic investment strategy
  employing a hybrid approach that combines LSTM and ARIMA models
  referred to as LSTM-ARIMA. This unique algorithm uses LSTM to produce
  final predictions but boost results of this RNN by adding the
  residuals obtained from ARIMA predictions among other inputs. The
  algorithm is tested across three equity indices (S\&P 500, FTSE 100,
  and CAC 40) using daily frequency data spanning from January, 2000 to
  August, 2023. The architecture of testing is based on the walk-forward
  procedure which is applied for hyperparameter tunning phase that uses
  using Random Search and backtesting the algorithms. The selection of
  the optimal model is determined based on adequately selected
  performance metrics combining focused on risk-adjusted return
  measures. We considered two strategies for each algorithm: Long-Only
  and Long-Short in order to present situation of two various groups of
  investors with different investment policy restrictions. For each
  strategy and equity index, we compute the performance metrics and
  visualize the equity curve to identify the best strategy with the
  highest modified information ratio (IR**). The findings conclude that
  the LSTM-ARIMA algorithm outperforms all the other algorithms across
  all the equity indices what confirms strong potential behind hybrid
  ML-TS (machine learning - time series) models in searching for the
  optimal algorithmic investment strategies.
  \end{abstract}
  
  
 \end{frontmatter}

\linespread{1.5}\selectfont

\hypertarget{introduction}{%
\section{Introduction}\label{introduction}}

Predicting the financial market is known to be quite challenging due to
factors such as volatility, the complexity of the financial system, and
the constantly changing economic landscape. We noticed twice in the past
20 years, the 2008 recession and the COVID-19 pandemic, that there was
so much uncertainty on how the markets will progress. Researchers and
traders try many approaches to successfully predict the financial
market. Unfortunately, not all are successful as it depends on the
economic and political situation of the stock markets. They try to
optimize their models ranging from simple linear regression models to
advanced machine learning (ML) algorithms and being tested based on all
types of invested assets.

High-frequency trading is gaining much more popularity; however, due to
its complexity, it may not be available for all users. Therefore, for
this research, we will consider only daily data. We believe that it can
still give us a general picture of the interactions in the market and
enable us to develop an algorithmic investment strategy (AIS). The main
focus of this study is to utilize the Auto-regressive Integrated Moving
Average (ARIMA) and Long-Short Term Memory (LSTM) models and combine
them into a hybrid model called LSTM-ARIMA. The ultimate goal is to
apply this hybrid model to develop an efficient AIS. Our main hypothesis
states that the LSTM-ARIMA model will outperform other algorithms in
most cases*. The ensuing are the research questions that our study aims
to explore:

\begin{quote}
\emph{RQ1. Are the algorithmic investment strategies robust to changes
in the asset?} \newline  \emph{RQ2. Does LSTM-ARIMA perform better than
the models individually?} \newline \emph{RQ3. Are the algorithmic
investment strategies robust to changes in the model hyperparameters?}
\newline \emph{RQ4. Does the Long-Only or Long-Short strategy outperform
the Buy\&Hold?}
\end{quote}

To evaluate our algorithmic investment strategies, we have selected
three assets, namely, S\&P 500 (\emph{GSPC}), FTSE 100 (\emph{FTSE}),
and CAC 40 (\emph{FCHI}) equity indices. The motivation behind the
choice of assets was to diversify its results across various equity
indices to capture the finest capability of the AIS. Therefore, each
asset is chosen from stock markets of different regions, from the New
York Stock Exchange (NYSE) to the London Stock Exchange (LSE) and
Euronext Stock Market (PAR). Our in-sample data begins on
\emph{2000-01-03} for S\&P 500 and CAC 40 and \emph{2000-01-04} for FTSE
100. The out-of-sample data for the S\&P 500 equity index starts on
2005-01-25, for the FTSE 100 equity index on 2005-01-13, and the CAC 40
equity index on 2004-12-28 and lasts until \emph{2023-08-30}. We capture
the horizon of approximately 23 years. During this time frame, we
capture two extreme times of the market. Considering extreme market
conditions while training the model can help them perform well during
both stable and volatile conditions.

Our contribution to the existing literature can be summarized in the
following sentences. Firstly, we develop algorithmic investment
strategies based on the predicted closing prices from ARIMA, LSTM, and
LSTM-ARIMA models and finally, we combine these forecasts into one
ensemble model additionally boosting its results. Additionally, we use
walk-forward optimization (WFO) as this technique reduces the risk of
over-fitting to one specific sample of past returns. For each walk, we
divide our data into training, validation, and testing data sets, where
the training set equals 1000 trading days and the validation and testing
set equals 250 trading days. Moreover, we perform hyperparameter tuning
at every walk by performing a random search using a set of parameters
explained later on in the study. Finally, the paper is finished with a
sensitivity analysis of the most promising model to verify its
robustness and potential for using it in real-time investments in
financial markets. From a broad literature review, we have concluded,
that there are very few papers that cover the process of testing
algorithmic investment strategies in such a complex and reliable way.

The structure of the paper is as follows: Section 1 contains an
introduction. Section 2 presents a brief overview of the literature.
Section 3 provides us with the data description. Section 4 defines the
methodology describing ARIMA, LSTM, and LSTM-ARIMA models. It also
presents the WFO, performance metrics, research description, and
hyperparameter tuning. Section 5 covers the empirical results of the
strategies using the equity curves and performance metrics. Section 6
presents the sensitivity analysis where we show the sensitivity of the
outcomes to changes in the set of hyperparameters. Section 7 presents
the ensembled AIS. The last section concludes and presents a further
extension of this paper.

\hypertarget{literature}{%
\section{Literature}\label{literature}}

Researchers are continuously searching for ways to build algorithmic
investment strategies (AIS) and make higher and less risky profits in
their investments This section will focus on the use of time series,
machine learning, and hybrid models to forecast the stock market prices
and create efficient algorithmic investment strategies.

\hypertarget{time-series-models}{%
\subsection{Time Series Models}\label{time-series-models}}

Time series models are considered to be well-performing as they can
catch the features of the financial time series data such as the
seasonality, trend, and cyclicality of historical data to predict the
future values. The time series model used in this study is ARIMA.
However, in this section, we will also discuss other types of models
used for time series analysis.

ARIMA, introduced by Box and Jenkins (1976), has been one of the main
tools of financial time series forecasting for a long time. ARIMA model
is derived from the ARMA model, by taking the first difference of the
prices to have a stationary data set. Mondal et al. (2014) studied the
effectiveness of ARIMA by forecasting fifty-six stocks from the Indian
stock market from different industries. For their predictions, they
achieved an accuracy of 85\% and the fast-moving consumer goods (FMCG)
sector was the most accurately predicted sector by ARIMA. Furthermore,
they also concluded that in ARIMA the change in training data size does
not influence the accuracy of their models.

Ariyo et al. (2014) used the ARIMA model to predict the prices of stocks
from the New York Stock Exchange (NYSE) and the Nigerian Stock Exchange
(NSE). They chose Nokia's and Zenith Bank stocks and the time frame was
16 years and 5 years respectively. The authors find the best model using
the Bayesian Information Criterion (BIC), the standard error of
regression, and the highest adjusted R-squared as the main criterion.
They concluded that the best model to predict Nokia's stock was
ARIMA(2,1,0) and for Zenith Bank, the best was ARIMA(1,0,1).

Devi et al. (2013) studied the effectiveness of ARIMA for the prediction
of stock trends. The authors selected the parameters based on manual
examination of ACF, and PACF plots to find the AR and MA orders. The
best model was selected based on the AIC and BIC criteria. The paper
considered five years of historical data for the analysis. The authors
conclude that ARIMA is the most accurate model to predict the stock
trend and make investment decisions.

Bui and Ślepaczuk (2022) explores the use of Hurst Exponent for an
algorithmic pair trading strategy. The authors also explored the use of
correlation and cointegration for their pair trading strategy. The study
is focused on 103 stocks from the NASDAQ 100 equity index, covering
approximately 18 years with daily frequency. The empirical findings
indicate that among all 103 stocks, the correlation method demonstrated
superior performance in terms of risk-adjusted return. However, the
Buy\&Hold strategy outperformed all other strategies in terms of
compounded annualized return.

Malladi and Dheeriya (2021) conducted a time series analysis of
cryptocurrency returns and volatility using GARCH, VAR, and ARMAX
models. ARMAX is an extension of the base ARMA which considers exogenous
inputs. They test the algorithm on BTC and XRP. The comparison is done
with a standard regression model. The conclusion is that both ARMAX and
GARCH perform better than the standard regression and the VAR model, as
expected. However, ARMAX showed the best results due to its high
accuracy.

Li et al. (2023) introduces spARIMA, a novel time series prediction
framework designed with a sequential training approach in batches. Named
for its sequential training based on noise levels and model fit
contributions, spARIMA incorporates a self-paced learning (SPL) strategy
to effectively mitigate data noise-induced instability. The model's
performance is evaluated across twelve diverse datasets, including
equity index prices (NASDAQ, RUSSEL, NYSE), as well as traffic and
temperature data. Comparisons are made with traditional ARIMA models
using two gradient descent algorithms. While spARIMA did not
consistently outperform ARIMA across all datasets, it demonstrates
promising capabilities against strong noise in time series prediction
tasks, demonstrating its potential in enhancing forecasting accuracy.

The Vector Autoregression (VAR) model is also known for its time series
modeling capabilities. Suharsono et al. (2017) uses VAR and VECM to
model the stock price. They use the ASEAN share price index and perform
a manual search to find the best parameters for the models. The criteria
to check the performance of the model was based on the Akaike
information criterion (AIC). They concluded that in comparison to the
VECM model, the VAR model performed the best in modeling.

Castellano Gómez and Ślepaczuk (2021) analyzed four algorithmic
strategies and one of them was based on ARIMA. They used S\&P 500 equity
index data for the predictions and used almost 31 years of historical
data. The goal was to create a portfolio strategy using four selected
algorithmic strategies. All the strategies were compared with the
benchmark buy\&hold. The paper showed that ARIMA did not perform well
when compared to the Buy\&Hold strategy, however, the performance of
ARIMA was the highest during the phases of high volatility.

\hypertarget{machine-learning-models}{%
\subsection{Machine Learning Models}\label{machine-learning-models}}

Machine learning is an advanced approach based on artificial
intelligence which can be used to forecast stock market prices. In
recent years, Recurrent Neural Networks (RNN) have started to be used
more often for time series analysis. Rumelhart (1986) made significant
contributions to the field of RNN. Due to issues such as the vanishing
gradient problem and the inability to effectively capture long-term
dependencies, the development of RNNs faces certain limitations.
Hochreiter and Schmidhuber (1997) proposed the architecture of the
Long-Short-Term Memory (LSTM) model to tackle the vanishing gradient
problem. This gives LSTM a huge advantage over RNN especially when it
comes to time series analysis.

Xiong et al. (2014) presents an innovative approach leveraging a firefly
algorithm (FA) to optimize multi-output support vector regression (MAVR)
parameters in financial forecasting. Their study evaluates this FA-MAVR
model across statistical, economic, and computational criteria.
Statistical evaluation includes goodness-of-forecast measures and
testing methodologies, while economic criteria assess the model's
performance using a naive trading strategy. Computational efficiency is
also considered. Testing is conducted on major equity indices: S\&P 500,
FTSE 100, and Nikkei 225. In comparison to genetic algorithms and
particle swarm optimization, FA-MAVR demonstrates superior forecast
accuracy and profitability, establishing its effectiveness in equity
indices price prediction.

Siami-Namini et al. (2018) compared the use of LSTM and ARIMA in
forecasting time series. They used them to predict the monthly closing
prices for eleven stock market indices. In comparison to ARIMA, they
conclude that LSTM outperforms the ARIMA model, which results in RMSE
measure lower by 85\%. Furthermore, they also mentioned that LSTM
results were robust to the number of epochs used in the process of
estimation.

Grudniewicz and Ślepaczuk (2023) researched various machine learning
techniques for creating an AIS. They utilized various machine learning
models, including Neural Networks, K Nearest Neighbours, Regression
Trees, Random RandomForest, Naive Bayes classifiers, Bayesian
Generalized Linear Models, and Support Vector Machines. These ML models
were employed to generate trading signals for WIG20, DAX, S\&P500, and
six CEE indices over a timeframe spanning approximately 21 years. The
authors concluded that in terms of risk-adjusted returns, the Polynomial
Support Vector Machine model performed the best in the case of WIG20 and
S\&P 500 equity indices, while the Linear Support Vector Machine model
for DAX and six CEE equity indices.

Roondiwala et al. (2017) presented a study predicting stock prices using
LSTM. Five years of historical data on the NIFTY 50 index was used for
testing purposes. The training of LSTM models was done by allocating
random weights and biases with an architecture of two LSTM layers and
two dense layers with ReLU and Linear activation function respectively.
Finally, the predicted values were compared with the actual values and
evaluated using the RMSE. The best RMSE score was given for the model
with High/Low/Open/Close as the inputs with 500 training epochs.

Michańków et al. (2022) presented a study on using LSTM in Algorithmic
Investment Strategies (AIS) on BTC and S\&P500 Index. The output of
their model was a singular value predicting the next day's return value
-1, 0, 1. The set of hyperparameters used for the tuning process,
relating to this paper, were the number of layers between 1 and 5, the
number of neurons in each layer chosen between 5 and 512, dropout rates
between 0.001 and 0.2, several types of optimizers including SGD,
RMSProp, and Adam variants, learning rates chosen between 0.001 and 0.1,
and the batch size ranging from 16 to the length of the test size. After
the hyperparameters tuning phase, they selected the model with 3 hidden
layers, with 512/256/128 neurons respectively, a dropout rate equal to
0.02, Adam as an optimizer a learning rate of 0.00015, and a batch size
of 80. They deduced that when it comes to daily frequency, their model
for S\&P 500 equity index performed well for the Long-Only strategy,
while the model for BTC performed well for both the Long-Only and
Long-Short strategy.

Zhang et al. (2019) wrote an analysis of the Attention-based LSTM model
for financial time series prediction (AT-LSTM). Instead of making their
prediction of LSTM by inputting the prediction of ARIMA, the authors use
the output of the attention model as the input of LSTM. The authors
compared the results of AT-LSTM with ARIMA and LSTM. The testing and
training were done on three data sets: Russell 2000, DIJA, and NASDAQ
indices, and the best model was concluded based on the MAPE (mean
absolute percentage error). LSTM performed the best with 2 layers, 8
hidden neurons, 20 training time steps, batch size equal to 50, and 5000
epochs. Finally, the authors summarized that the hybrid AT-LSTM
performed better than LSTM and both of them performed way better than
ARIMA.

Baranochnikov and Ślepaczuk (2022) analyzed various architectures of
LSTM and GRU models in Algorithmic Investment Strategies. Their LSTM
model forecasted the rate of return for the period T+1. The authors
decided to use the set of parameters chosen from the financial
literature. Ten model architectures were used during the training
process and parameters such as dropout rate, batch size, epochs, and the
learning rates were additionally modified. Adam optimizer with the
AMSGrad extension was used in all. The authors used the walk-forward
process for estimation purposes. The models were tested on Bitcoin,
Tesla, Brent Oil, and Gold closing prices. The authors deduced that the
LSTM outperformed the traditional Buy\&Hold strategy for Bitcoin and
Tesla both for daily and hourly frequency.

Another research was conducted in India by Hiransha et al. (2018) who
predicted the National Stock Market of India and the New York Stock
Exchange using various algorithms such as MLP, RNN, LSTM, and CNN. ARIMA
was used as a benchmark. The authors tested three sectors of industries,
automobile, finance, and IT from both stock exchanges and measured the
accuracy of the predictions using the MAPE metric. The authors concluded
that CNN (Convolution Neural Network) outperformed all the other models
and that LSTM was better performing than ARIMA due to its useful
capability of finding non-linearity.

Kijewski and Ślepaczuk (2020) predicted the S\&P 500 equity index prices
using the classical models and RNN (Recurrent Neural Networks including
ARIMA, MA, momentum and contrarian, volatility breakout, macro factor,
and finally LSTM. The models were trained and tested collectively over
twenty years. The range of parameters taken for the ARIMA model were, p:
0-5, d: 0-3, and q: 0-5, while in the case of LSTM, a prepared set of
hyperparameters was taken from the literature. They mentioned that the
best LSTM model had the following configuration: 30 neurons in the
hidden layers with ReLU activation, length of sequence equaling 15, and
dropout rate 0.02 by using Adam optimizer with learning rate 0.01 and
loss function Mean Square Error. The authors concluded that LSTM
outperformed ARIMA and the benchmark buy\&hold strategy.

Kryńska and Ślepaczuk (2022) tested several architectures of the LSTM
model in AIS based on the S\&P 500 equity index and BTC. They tested
three frequencies of data: daily, hourly, and 15-minute of S\&P 500 and
BTC. When the model was considering a regression problem, models on
daily data performed better than intraday frequencies. However, in the
case of classification problems, the model on intraday data performed
the best.

Nelson et al. (2017) used LSTM to predict the stock market direction.
The authors use different stocks from the Brazilian stock exchange and
some technical indicators as inputs. Furthermore, log-return
transformation was performed for the inputs and the frequency of data
was 15 minutes. The output of the model is a binary value (1, 0)
denoting an increase and decrease in the prices between the time steps.
Four metrics were created to evaluate the performance, among which were
the accuracy and the precision. The authors concluded that the proposed
model of the paper outperforms the benchmark Buy\&Hold strategy based on
accuracy and offers less risky investment compared to the others.

Mizdrakovic et al. (2024) investigates Bitcoin price dynamics by
analyzing factors including Ethereum, S\&P 500, VIX, EUR/USD, and
GBP/USD. The study introduces a two-tiered methodology: initially
employing variational mode decomposition (VMD) enhanced with a variant
of the sine cosine algorithm to optimize VMD's control parameters for
trend extraction from time series data. Subsequently, LSTM and hybrid
Bidirectional LSTM models are utilized to forecast prices over multiple
time steps. The authors evaluate their approach across various feature
sets, comparing performances with and without VMD. Their findings
highlight the outperformance of the V-BiLSTM-HSA-SCA (VMD Bidirectional
LSTM with hybrid self-adaptive sine cosine algorithm) model,
demonstrating its highest R2 and IA scores, as well as lowest MAE, MSE,
and RMSE values among all tested models.

\hypertarget{hybrid-models}{%
\subsection{Hybrid Models}\label{hybrid-models}}

Hybrid models are a combination of two models. A mix of models may work
better together as they capture the efficiencies of individual models.
For instance, LSTM-ARIMA models can help us capture the linear and
non-linear dependencies in the data. Most algorithms can be combined.
One of the common combinations is a mixture of ARIMA and GARCH models.
Vo and Ślepaczuk (2022) tested a hybrid ARIMA-SGARCH model in
algorithmic investment strategies (AIS). Three models were created
during their research: ARIMA, ARIMA-SGARCH, and ARIMA-EGARCH. The models
were tested on the S\&P 500 equity index and prices in the period of the
last 20 years. The authors selected the best ARIMA model using the
Akaike Information Criterion (AIC). The results were compared between
the models and the benchmark buy\&hold strategy. Several performance
metrics were used. The authors concluded that the ARIMA-SGARCH model
performed the best, followed by ARIMA which performed better than the
benchmark.

Senneset and Gultvedt (2020) uses ARIMA and LSTM together to increase
their portfolio stability. They used several stocks from the Oslo Stock
Exchange for 14 years. The authors used the residual values from ARIMA
as an input and the performance was compared using the RMSE and MAE
error metrics. During the research, two hybrid models were created:
ARIMA-RandomForest and ARIMA-LSTM. The results concluded that
ARIMA-RandomForest outperformed all the strategies, while ARIMA-LSTM
outperformed just the benchmark strategy.

LSTM-ARIMA was also considered in the approach to forecasting the wind
speed by Bali et al. (2020). A few wind parameters such as wind speed,
temperature, pressure, etc were used as inputs for LSTM. The authors
compared the LSTM-ARIMA model with LSTM and the support vector machine
(SVM) using the RMSE. They concluded that the LSTM-ARIMA was the most
accurate model compared to LSTM and SVM.

Arnob et al. (2019) forecasted the Dhaka stock exchange (DSE), using the
hybrid ARIMA-LSTM approach. The aim was to forecast the correlation
coefficient between the assets. They used fifteen companies from DSE to
forecast. Several ARIMA orders were chosen using the ACF and PACF plots;
however, the best order was chosen based on the lowest AIC. The data was
divided into three parts: train, test1, and test2, and the performance
was measured using MSE and MAE. The researchers concluded that
ARIMA-LSTM performed better than the ARIMA model.

Karim et al. (2022) predicted the stock price of NIFTY-50 stock using a
bidirectional LSTM and GRU network hybrid model (Bi-LSTM-GRU). The
hybrid model was compared with each model being trained individually and
the one with the highest precision was marked as the best. The authors
concluded that their proposed hybrid approach outperformed the models
individually.

Oyewola et al. (2024) present an approach to stock price prediction
within the oil and gas sector, an industry with complex market dynamics
and diverse external influences. The study introduces three models: Deep
Long Short-Term Memory Q-Learning (DLQL), Deep Long Short-Term Memory
Attention Q-Learning (DLAQL), and the state-of-the-art Long Short-Term
Memory (LSTM) model. Historical stock prices of CVE, MPLX LP, LNG, and
SU are utilized for training and evaluation. The reinforcement learning
technique employed is the Markov Decision Process (MDP) framework.
Results indicate that the DLAQL model outperforms all the others in
decision-making capabilities, risk management, and most importantly, the
profitability, positioning it as a robust choice for stock price
prediction in the oil and gas sector.

To summarize, the ARIMA model has been extensively used for financial
time series forecasting. Studies show this model's capabilities in
predicting stock prices from various exchanges. The results vary based
on the statistical criteria chosen such as AIC and BIC. Other methods
were also explored such as Vector Autoregression (VAR) or Vector Error
Correction Models (VECM) for stock price modeling purposes. Furthermore,
deep learning techniques such as the Long Short-Term Memory model
(LSTM), have gained traction for time series forecasting. LSTM is a
model designed to address issues like vanishing gradients in traditional
Recurrent Neural Networks (RNN) and shows superior performance when
compared to ARIMA. Other deep learning techniques, including Support
Vector Machines (SVM) or Random Forest (RF), have also been investigated
and demonstrated improved predictive capabilities compared to
traditional models like ARIMA.

Based on the summary provided, we propose that combining the two
top-performing models, namely ARIMA and LSTM, could help overcome the
limitations of each model. ARIMA excels at capturing short-term time
series patterns, while LSTM is adept at modeling long-term dependencies.
By integrating these models, we anticipate outperforming their
capabilities and potentially gaining an edge in beating the market.

\hypertarget{data-description}{%
\section{Data Description}\label{data-description}}

We consider three equity indices in our research and the data is taken
from Yahoo Finance using their \emph{yfinance} API for \emph{Python}.
Table 1 presents the descriptive statistics of the chosen assets. Please
note that the difference in the count value may arise due to the
difference in the number of trading days in a year in each country.

\renewcommand{\arraystretch}{0.75}
\fontsize{9}{9.5}\selectfont

\begin{longtable}[]{@{}
  >{\centering\arraybackslash}p{(\columnwidth - 16\tabcolsep) * \real{0.1370}}
  >{\centering\arraybackslash}p{(\columnwidth - 16\tabcolsep) * \real{0.0959}}
  >{\centering\arraybackslash}p{(\columnwidth - 16\tabcolsep) * \real{0.0822}}
  >{\centering\arraybackslash}p{(\columnwidth - 16\tabcolsep) * \real{0.2740}}
  >{\centering\arraybackslash}p{(\columnwidth - 16\tabcolsep) * \real{0.0822}}
  >{\centering\arraybackslash}p{(\columnwidth - 16\tabcolsep) * \real{0.0822}}
  >{\centering\arraybackslash}p{(\columnwidth - 16\tabcolsep) * \real{0.0822}}
  >{\centering\arraybackslash}p{(\columnwidth - 16\tabcolsep) * \real{0.0822}}
  >{\centering\arraybackslash}p{(\columnwidth - 16\tabcolsep) * \real{0.0822}}@{}}
\caption{Descriptive Statistics for the closing price
series.}\tabularnewline
\toprule\noalign{}
\begin{minipage}[b]{\linewidth}\centering
\end{minipage} & \begin{minipage}[b]{\linewidth}\centering
Count
\end{minipage} & \begin{minipage}[b]{\linewidth}\centering
Mean
\end{minipage} & \begin{minipage}[b]{\linewidth}\centering
Standard Deviation
\end{minipage} & \begin{minipage}[b]{\linewidth}\centering
Min
\end{minipage} & \begin{minipage}[b]{\linewidth}\centering
25\%
\end{minipage} & \begin{minipage}[b]{\linewidth}\centering
50\%
\end{minipage} & \begin{minipage}[b]{\linewidth}\centering
75\%
\end{minipage} & \begin{minipage}[b]{\linewidth}\centering
Max
\end{minipage} \\
\midrule\noalign{}
\endfirsthead
\toprule\noalign{}
\begin{minipage}[b]{\linewidth}\centering
\end{minipage} & \begin{minipage}[b]{\linewidth}\centering
Count
\end{minipage} & \begin{minipage}[b]{\linewidth}\centering
Mean
\end{minipage} & \begin{minipage}[b]{\linewidth}\centering
Standard Deviation
\end{minipage} & \begin{minipage}[b]{\linewidth}\centering
Min
\end{minipage} & \begin{minipage}[b]{\linewidth}\centering
25\%
\end{minipage} & \begin{minipage}[b]{\linewidth}\centering
50\%
\end{minipage} & \begin{minipage}[b]{\linewidth}\centering
75\%
\end{minipage} & \begin{minipage}[b]{\linewidth}\centering
Max
\end{minipage} \\
\midrule\noalign{}
\endhead
\bottomrule\noalign{}
\endlastfoot
S\&P 500 & 5953 & 1939 & 1027 & 677 & 1190 & 1449 & 2486 & 4797 \\
FTSE 100 & 5975 & 6057 & 1050 & 3287 & 5332 & 6150 & 6867 & 8014 \\
CAC 40 & 6049 & 4714 & 1118 & 2403 & 3798 & 4570 & 5476 & 7577 \\
\end{longtable}

\normalsize
\vspace{-0.3cm}

\noindent\linespread{0.65}\selectfont {\scriptsize \textbf{Note}: \textit{The descriptive statistics for S\&P 500, FTSE 100, and CAC 40 are calculated on the closing price in the period from 2000-01-03 for S\&P 500 and CAC 40 and 2000-01-04 for FTSE 100 until 2023-08-31.}}

\linespread{1.5}\selectfont

\pagebreak

Figure 1 presents the prices of tested assets and their volatilities.
For the S\&P 500 equity index, we use the CBOE's VIX as the measure of
volatility and for the FTSE 100 and CAC 40 equity index we use the
annualized realized volatility with a historical window of 21 days,
which is calculated using the following formula:

\[
annRV =  \sqrt{\frac{\sum_{i=21}^{N} R_t^2 }{N}} \times \sqrt{252} \tag{1}
\] where: \newline \(R_t\) - the daily returns \newline \(t\) - the
counter representing each trading day \newline \(N\) - the number of
trading days in our time frame

\begin{figure}
\centering
\includegraphics[width=0.9\textwidth]{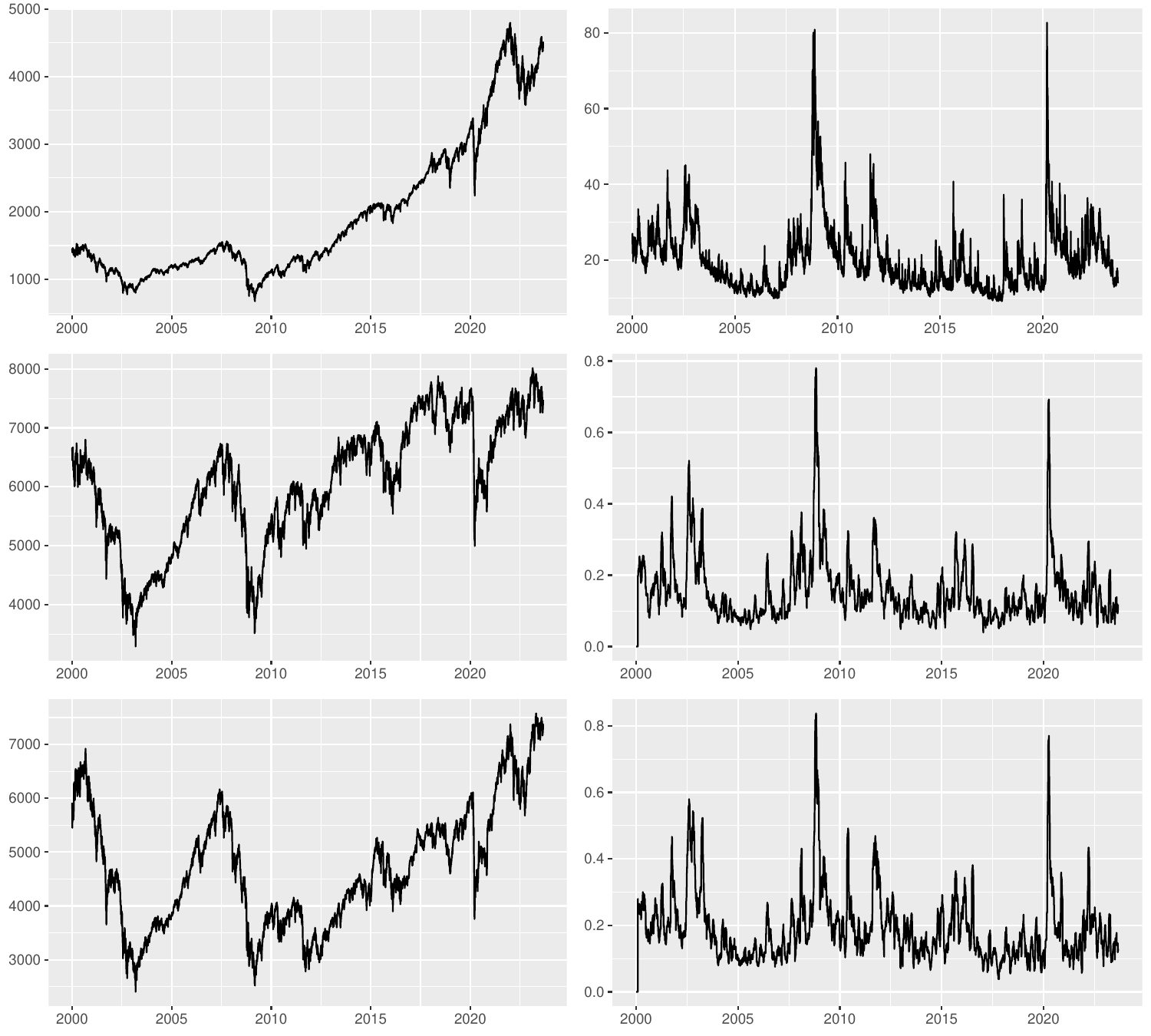}
\caption{Index Prices and their Volatilities.}
\end{figure}

\vspace{-0.5cm}

\noindent\linespread{0.65}\selectfont{\scriptsize\textbf{Note}: \textit{Each plot on the left shows the behavior of the index based on their closing price and each plot on the right represents the volatility for the corresponding index. Note that for S\&P 500, CBOE's VIX has been used as a measure of volatility, and for the FTSE 100 and CAC 40, the realized volatility was used with a window of 21 trading days.}}

\linespread{1.5}\selectfont

\hypertarget{methodology}{%
\section{Methodology}\label{methodology}}

\hypertarget{arima}{%
\subsection{ARIMA}\label{arima}}

Autoregressive Integrated Moving Average model (ARIMA), is an
econometric model used for forecasting time series data based on some
historical data. The model was introduced by George Box and Gwilym
Jenkins in 1976, and they initially used it to model changes in
financial time series data. The model consists of three parts:
Autoregressive (AR), Integrated (I), and Moving Average (MA), where each
component has its order. Let's denote \emph{p}, an order for \emph{AR}
component, \emph{d}, an order for \emph{I} component, and \emph{q}, an
order for \emph{MA} component. Denoting this we can then write the model
as \emph{ARIMA(p,d,q)}. The respective orders determine the following
properties of the model:

\begin{itemize}
\tightlist
\item
  \emph{p} - the number of lagged observations
\item
  \emph{d} - the number of times the data was differenced
\item
  \emph{q} - the order of the MA process
\end{itemize}

In terms of stock forecasting, order \emph{d} is usually set to 1 when
we model the prices and 0 when we model the returns as the data is
already stationary, and then we have \emph{ARMA(p,q)}. Figure 2 presents
the difference between the non-stationary data set which is the closing
prices and the stationary data set which are the first differences of
prices.

\begin{figure}
\centering
\includegraphics{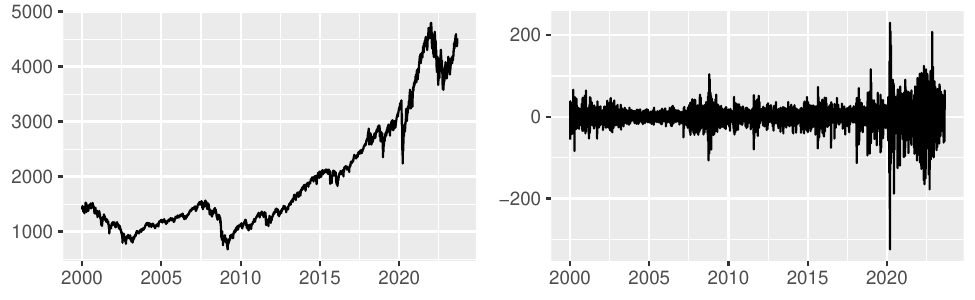}
\caption{Non-stationary data set compared to stationary data set of S\&P
500.}
\end{figure}

\vspace{-0.5cm}

\noindent\linespread{0.65}\selectfont {\scriptsize \textbf{Note}: \textit{The plot on the left side shows an actual plot of S\&P 500 closing price and the plot on the left side shows the differentiated closing price.}}

\linespread{1.5}\selectfont

Let's now look at the formal side of an ARIMA model. ARIMA generally is
an extension of the ARMA model. The \emph{AR(p)} can be denoted with the
following equation: \[
AR(p): y_t = \phi_1 y_{t-1} + \phi_2 y_{t-2} + ... + \phi_p y_{t-p} + \epsilon_t \tag{2}
\] where: \newline \(y_t\) - the value of the time series at a time
\emph{t} \newline \(\epsilon_t\) - the error term \newline \(\phi\) -
the coefficients that capture the relationship between the current
observation and previous observations at a lag of \emph{p} \newline The autoregressive component is responsible for forecasting the chosen
variable using the past value of the variable automatically. \newline The \emph{AR(p)} model can be written as: \[
AR(p): (1-\sum_{i=1}^{p}\phi_iL^i)y_t = c + \epsilon_t \tag{3}
\] The second component that is responsible for differencing can be
denoted as \emph{I(d)} and presented as: \[
I(d): (1-L)^d = \mu + \epsilon_t \tag{4}
\] The third component, moving average, looks as follows: \[
MA(q): y_t =  \mu + \epsilon_t + \theta_1 \epsilon_{t-1} + ... + \theta_q \epsilon_{t-q} \tag{5}
\] where: \newline \(\mu\) - the mean of the given series \newline
\(\theta_1\space...\space\theta_q\) - the respective weights for each
error term \(\epsilon_{t-1}\space...\space\epsilon_{t-q}\). \newline This represents the moving average procedure with order \emph{q}. Unlike
\emph{AR(p)}, the \emph{MA(q)} uses the previous error terms for the
regression. And using the lag operator, \emph{MA(q)} may be denoted as:
\[
MA(q): y_t = \mu + (1+\sum_{i=1}^{q}\theta_iL^i)\epsilon_t \tag{6}
\] Using the two components explained previously, we can find the
\emph{ARIMA(p,0,q)} model which is also known as \emph{ARMA(p,q)} model.
The equation of \emph{ARMA(p,q)} looks as follows: \[
ARMA(p,q): y_t = \phi_1y_{t-1} + \phi_2y_{t-} + ... + \phi_py_{t-p} + \epsilon_t - \theta_1\epsilon_{t-1} + ... - \theta_q\epsilon_{t-q} \tag{7}
\] The final \emph{ARIMA(p,d,q)} can be written as: \[
ARIMA(p,d,q): (1-\sum_{i=1}^{p}\phi_iL^i)(1-L)^dy_t = c + (1+\sum_{i=1}^{q}\theta_iL^i)\epsilon_t \tag{8}
\] where: \newline \(d\) - the number of times the series was
differenced.

\hypertarget{rnn}{%
\subsection{RNN}\label{rnn}}

According to Turing (2023), a job platform, a recurrent neural network
(RNN) is a variation of artificial neural networks (ANN). RNN may be
used to address various problems such as speech recognition or image
captioning. What differentiates RNN from ANN is that ANN just takes the
inputs and generates outputs; however, RNN learns from the previously
generated outputs to provide results for the next time stamp. Another
advantage of RNN is that it has a memory cell that continues the
calculations and if the forecast is inaccurate the network auto-learns
and executes backpropagation to get the correct result. RNN is very
effective for time series forecasting due to its ability to recollect
previous inputs. This is where the Long-Short-term memory (LSTM) model
comes in.

\hypertarget{lstm}{%
\subsubsection{LSTM}\label{lstm}}

LSTM is a special type of RNN. Its main character is the ability to
handle long-term data dependencies and push the outcome to the
succeeding node more efficiently. It also addresses the vanishing
gradient problem, a known issue with RNN, which is tackled by
disregarding nugatory information using its forget gate. LSTM also deals
well with long-term dependencies i.e.~with problems where the output is
dependent on the historical inputs. LSTM consists of multiple gates,
each having an essential task to be done to have positive results.

Figure 3 presents LSTM which has four gates: input, output, forget, and
change. For a sequence in time \(x_t - (x_1, x_2,...,x_n )\) the forget
gate \(f_t\) takes the \(x_t\) and the hidden state \(h_{t-1}\) and
produces a binary output 0 and 1 through a sigmoid function and
identifies which information should be discarded from the memory cell
\(c_{t-1}\). The value equal to 1 is forwarded to the cell with the
value equal to 0 and all the other information is forgotten. The input
gate \(i_t\) identifies what to update from the change gate
\(\hat{c}_{t}\) and the output gate \(o_t\) decides which information
should be taken from the current cell. From the sequence \(X\), two
sequences \(x\) and \(y\) are created, where \(x\) is the input sequence
and \(y\) is the next day closing price. Furthermore, it's worth noting
that the memory cell is responsible for long-term memory and it updates
the input gate, forget gate and the change gate. On the other hand, the
hidden state is responsible for the short-term memory and is updated by
the output gate and the memory cell. The explanation above was
influenced by Bhandari et al. (2022) research on `Predicting stock
market index using LSTM'.

\begin{figure}

{\centering \includegraphics[width=0.7\linewidth]{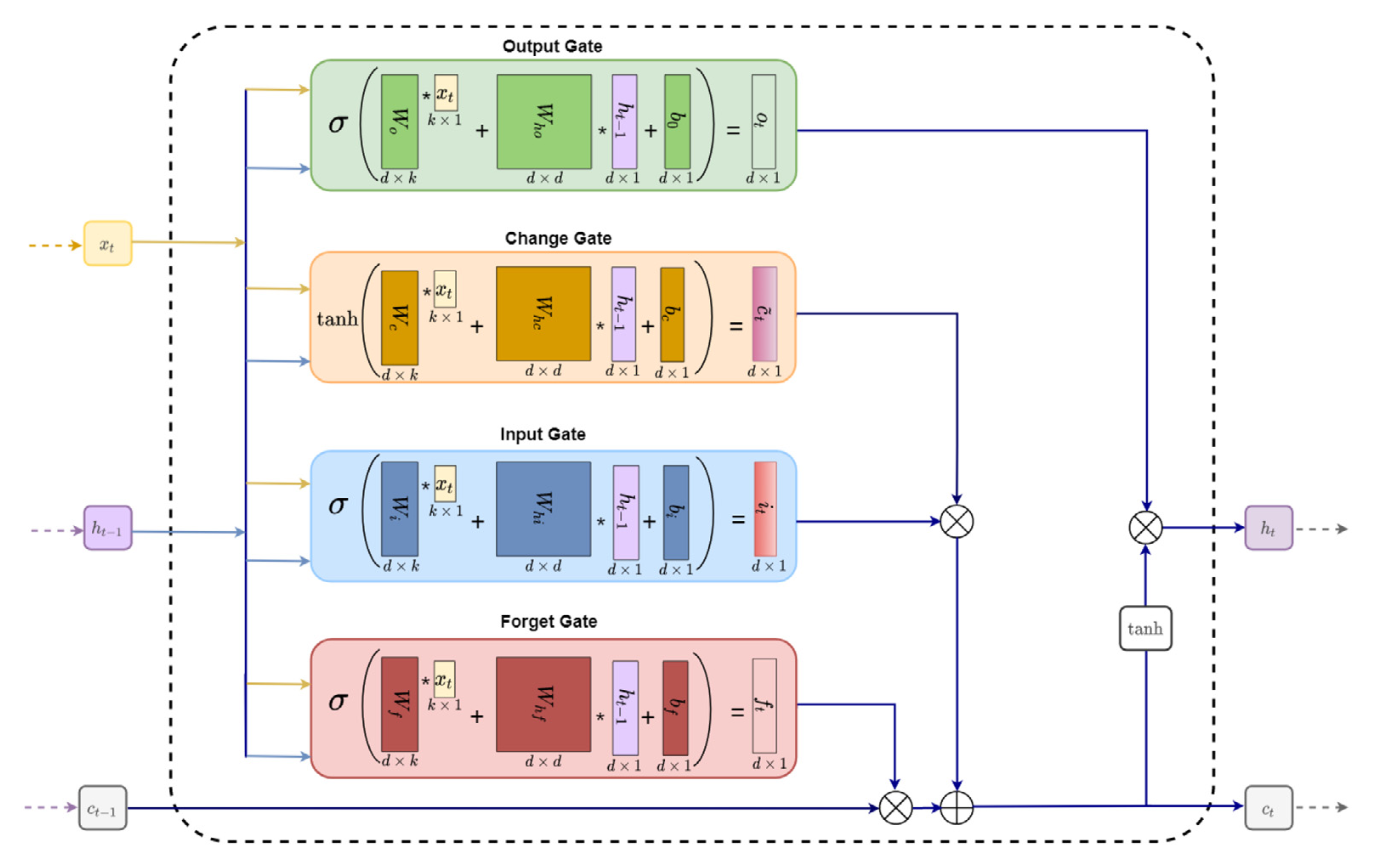} 

}

\caption{The architecture of Long-Short-term memory.}\label{fig:LSTM_}
\end{figure}
\begin{center}
\vspace{-0.4cm}\noindent\linespread{0.65}\selectfont {\scriptsize \textbf{Note}: \textit{The architecture of LSTM, source: https://www.sciencedirect.com/science/article/pii/S2666827022000378.}}
\end{center}
\linespread{1.5}\selectfont 

The mathematical equations to the previously given terminologies look as
follows: \[
i_t = \sigma(W_ix_t\space+\space W_{hi}h_{t-1}\space+\space b_i) \tag{9}
\] \[
f_t = \sigma(W_fx_t\space+\space W_{hf}h_{t-1}\space+\space b_f) \tag{10}
\] \[
o_t = \sigma(W_ox_t\space+\space W_{ho}h_{t-1}\space+\space b_o) \tag{11}
\] \[
\hat{c}_t = tanh(W_cx_t\space+\space W_{hx}h_{t-1}\space+\space b_c) \tag{12}
\] \[
c_t = f_t \cdot c_{t-1} \space+\space i_t \cdot \hat{c}_t \tag{13}
\] \[
h_t = o_t \cdot tanh(c_t) \tag{14}
\] where: \newline \(W\) - weights \newline \(b\) - biases \newline
\(x_t\) - sequence of time \(t\) \newline \(f_t\) - forget gate at time
\(t\) \newline \(h_{t-1}\) - hidden state at time \(t-1\) \newline
\(i_t\) - input gate at time \(t\) \newline \(\hat{c}_{t}\) - change
gate at time \(t\) \newline \(o_t\) - output gate at time \(t\)

\hypertarget{lstm-arima}{%
\subsection{LSTM-ARIMA}\label{lstm-arima}}

In this paper, we introduce a hybrid approach using the ARIMA and LSTM
models collectively. This model contains an LSTM input layer which is
fed with the residuals of ARIMA predictions and other inputs such as the
closing price and the volatility. LSTM-ARIMA is a combination that helps
capture both the linear and non-linear properties of the data. Moreover,
LSTM is known for its outstanding capability to capture the long-term
dependencies in time series data and ARIMA is known for its outstanding
capability to capture the short-term dependencies in time series data.
Additionally, ARIMA learns from data using statistical methods and LSTM
learns by looking at the pattern thanks to the neural networks.
Considering all of the above strengths and weaknesses of the models, we
believe that collectively they may outperform the performance of both
individually. Generally, the process of using the LSTM-ARIMA approach
for AIS in this paper can be summarized in the following way:

\begin{enumerate}
\def\labelenumi{\arabic{enumi}.}
\tightlist
\item
  Find the best-fitted ARIMA model using the set of orders based on the
  smallest Akaike information criterion (AIC).
\item
  Get the residuals from ARIMA.
\item
  Perform feature engineering on LSTM, taking into consideration the
  residuals by ARIMA, the closing price, and the realized volatility of
  the asset under consideration (in the case of the S\&P 500, take VIX).
\item
  Conduct a random search and choose the \emph{best set of
  hyperparameters} based on the criteria outlined in the subsequent
  sections.
\item
  Fit the best model and execute predictions for buy/sell signals
  generations.
\item
  Create the equity curve based on investment signals from the previous
  point and then compute the performance metrics.
\end{enumerate}

\hypertarget{walk-forward-optimization}{%
\subsection{Walk Forward Optimization}\label{walk-forward-optimization}}

Over-fitting is a big risk in machine learning algorithms, especially in
financial time series forecasting. Common cross-validation techniques
like \emph{k-fold} are not well suited for financial analysis and
adjusting the hyperparameters may result in over-fitting. Common
cross-validation techniques sometimes do not perform as well as
intended. Therefore, to have a robust trading strategy it is advised to
use the walk-forward optimization (WFO) approach. Carta et al.~(2021)
stated that walk-forward optimization is one of the most popular
validation techniques used by financial researchers to undergo
decision-making for trading. There are two types of WFO: anchored, and
non-anchored. The difference lies that in anchored WFO each walk has a
common beginning point; however, in the non-anchored type each walk has
a different starting point but the same length. In this research, we
considered the non-anchored type for training, validation, and testing
as we believe its robustness is higher than the anchored type. We set
the in-sample (IS) window to 1250 trading days where the training set is
equal to 1000 trading and the validation set is equal to 250 trading
days and we set the out-of-sample (OOS) window to 250 trading days. This
is visualized in Figure 4.

\begin{figure}

{\centering \includegraphics[width=0.95\linewidth,height=0.6\textheight]{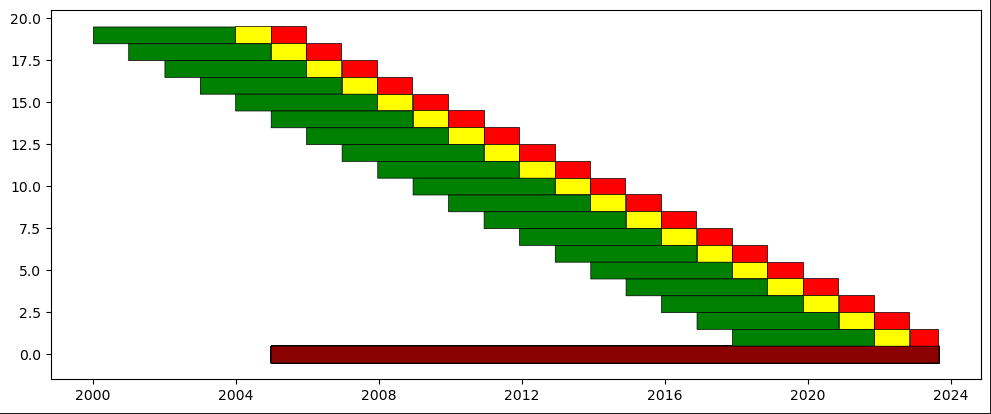} 

}

\caption{Walk forward optimization scheme with 5-years IS and 1-year OOS.}\label{fig:WFO}
\end{figure}
\vspace{-0.4cm}

\noindent\linespread{0.65}\selectfont {\scriptsize \textbf{Note}: \textit{The bars in green color represent the training data set, in yellow color represent the validation data set, in red color represent the out-of-sample testing data set, and the bars in the dark-red color represent the total out-of-sample data. This plot was designed by using the data for the S\&P 500 equity index. However, it looks similar for the FTSE 100 and CAC 40 equity indices. The training window is 1000 trading days, and validation and testing windows are 250 trading days each.}}

\linespread{1.5}\selectfont

\pagebreak

\hypertarget{performance-metrics}{%
\subsection{Performance Metrics}\label{performance-metrics}}

To assess and evaluate the robustness of the trading strategies created
in this paper, we calculated the performance metrics based on Michańków
et al.~(2022) and Bui and Ślepaczuk (2021). The details are presented:

Annualized Return Compounded \((ARC)\), shows the rate of return that
was annualized for the given strategy during the period of
\((0,...,T)\). It is expressed in percentage. \[
ARC = (\space\prod_{t=1}^{N} \space (1+R_t)\space)^{\frac{252}{N}} - 1 \times 100\% \tag{15}
\] where: \newline \(R_t\) - the percentage rate of return \newline
\(N\) - the sample size \newline \(R_t\) is calculated in the following way: \[
R_t = \frac{P_t - P_{t-1}}{P_{t-1}} \tag{16}
\] where: \newline \(P_t\) - the price at point \(t\) \newline

Annualized Standard Deviation \((ASD)\) is a risk measure. \[
ASD = \sqrt{252} \space\times\space \sqrt{\frac{1}{N-1}\sum_{t=1}^{N}\space(R_t\space - \bar{R})^2} \times 100\% \tag{17}
\] where: \newline \(R_t\) - the percentage rate of return \newline
\(\bar{R}\) - the mean rate of return \newline \(N\) - the sample size \newline \(\bar{R}\) is calculated in the following way: \[
\bar{R} = \frac{1}{N} \sum_{t=1}^{N} R_t \tag{18}
\]

Maximum Drawdown \((MD)\) gives us the maximum percentage drawdown
throughout the investment and is calculated as follows: \[
MD(T) = \mathop{\mathrm{max}}_{\tau\in[0,T]} \space(\mathop{\mathrm{max}}_{t\in[0,\tau]}(R_{i,T}\space-\space R_{i,\tau})) \times 100\% \tag{19}
\]

Maximum Loss Duration \((MLD)\) tells us about ``the number of years
between the previous local maximum to the forthcoming local maximum''
(Michańków et al.~(2022)) and is calculated as follows: \[
MLD = max(\frac{m_j\space-\space m_i}{S}) \tag{20}
\]

Information Ration \((IR^{*})\) describes the risk-adjusted return
metric based on the relation between ARC to its ASD and is calculated as
follows: \[
IR^{*} = \frac{ARC}{ASD} \times 100\% \tag{21}
\]

Modified Information Ratio \((IR^{**})\) is another more complex and
comprehensive risk-adjusted return metric which we regard as the
\textbf{most important} metric for the evaluation of strategies in this
research and is calculated as follows: \[
IR^{**} = IR^{*}\space\times\space ARC\space\times\space\frac{sign(ARC)}{MD} \% \tag{22}
\]

\hypertarget{research-description}{%
\subsection{Research Description}\label{research-description}}

In this study, we use a random search for hyperparameter tuning
conducted at each walk of WFO. The steps of the research are presented
below:

\begin{enumerate}
\def\labelenumi{\arabic{enumi}.}
\tightlist
\item
  Select the asset and download the data for 1-day frequency using the
  \emph{yfinance} Python API.
\item
  Perform data cleansing and prepare the data for feature engineering.
\item
  Create a code that supports the whole study including the sensitivity
  analysis.
\item
  Select the base model scenario for each model.
\item
  Run a random search and test the strategies.
\item
  Generate a prediction and take a position based on the criteria
  presented in the next section.
\item
  Create equity curves and calculate the performance metrics.
\item
  Conduct a sensitivity analysis.
\item
  Summarize the study with the best parameters.
\end{enumerate}

\pagebreak

In Diagram 1, the steps can be visually represented to provide a clearer
understanding.

\begin{center}
\noindent{\normalsize {Diagram 1}: {Research Description Flow Chart.}}
\vspace{-0.3cm}

\begin{center}\includegraphics[width=0.95\linewidth,height=0.8\textheight]{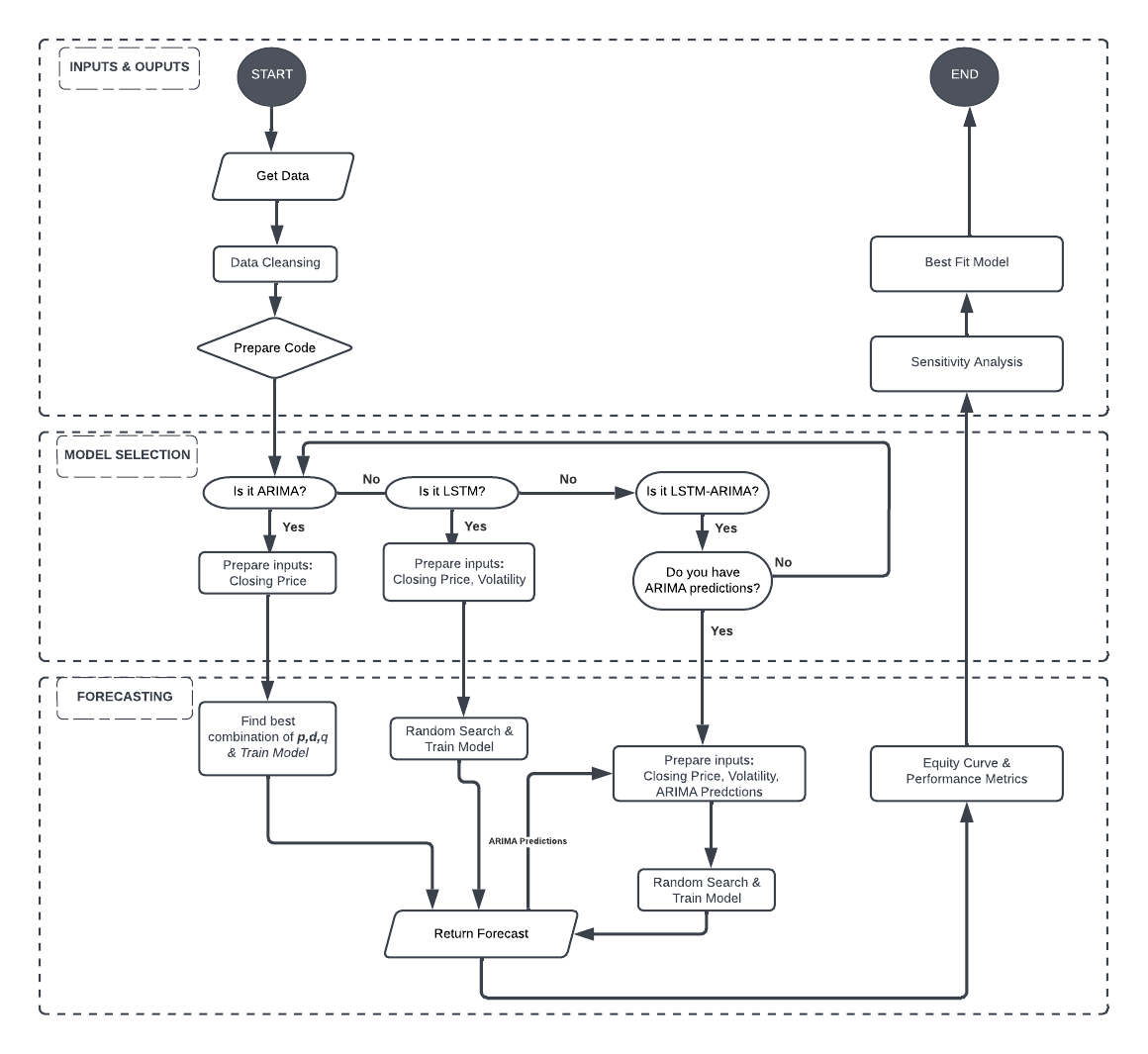} \end{center}

\vspace{-0.8cm}\noindent\linespread{0.65}\selectfont {\scriptsize \textbf{Note}: \textit{The diagram presents the process of research done in this study.}}
\end{center}

\linespread{1.5}\selectfont

\hypertarget{best-set-of-hyperparameters-criteria}{%
\subsection{Best Set of Hyperparameters
Criteria}\label{best-set-of-hyperparameters-criteria}}

In our research, we employ random search as a method of hyperparameter
tuning. During random search, we select five models with the lowest
validation loss. Then we calculate the \(IR2\) on the training data set
and the validation data set and then calculate the absolute value of
their difference. The \emph{best model} is the one with the lowest
absolute value of the difference and where the \(IR2\) for the
validation data set was \textbf{NOT} equal to \textbf{zero}.

\pagebreak

\hypertarget{strategy}{%
\subsection{Strategy}\label{strategy}}

During this research, we considered two kinds of strategies to be
evaluated: \emph{Long-Only} and \emph{Long-Short}. \emph{Long-Only} is
where we allow to open either a long position (1) or hold no position
(0). \emph{Long-Short} is where we allow to open either a long position
(1) or a short position (-1). The change for a Long position in both
strategies happens whenever the predicted price of \(t+1\) is higher
than the price at time \(t\) and we take a Short position or hold no
position whenever the predicted price of \(t+1\) is lower than the price
at time \(t\). Note that at this step we already have the prediction
generated and we use them to test the algorithm. Below we present the
mathematical notation of the Long-Only and Long-Short strategies: \[
Long-Only: \begin{cases}  
Signal = 1 \text{   if   } P_{t+1} > P_{t} \\
 Signal = 0 \text{   if   } P_{t+1} < P_{t}
\end{cases} 
\tag{23}
\]

\[
Long-Short: \begin{cases}  
Signal = 1 \text{   if   } P_{t+1} > P_{t} \\
 Signal = -1 \text{   if   } P_{t+1} < P_{t}
\end{cases} 
\tag{24}
\] where: \newline \(P_{t+1}\) - the closing price at \(t+1\) \newline
\(P_{t}\) - the closing prices at \(t\) \newline

\hypertarget{hyperparameter-tunning}{%
\subsection{Hyperparameter Tunning}\label{hyperparameter-tunning}}

In this section, we present to you the set of parameters that we use to
employ random search. Out of all combinations we conduct 20 trials on a
randomly chosen set. We perform the experiments in \emph{python 3.8.15}
using the \emph{Tensorflow} library. For ARIMA, the whole process from
model training, including all WFO walks, to generating the predictions
for a single asset took us approximately 3 minutes, for LSTM
approximately 3 hours, and for LSTM-ARIMA approximately 4 hours. Note
that for all models, the IS window is equal to 1250 trading days and the
OOS window is equal to 250 trading days. By default, we have set the
number of epochs to 100. However, using the \emph{Keras EarlyStop}
function we optimize the number of epochs based on the validation loss
while setting the patience to 10 epochs.

\hypertarget{arima-1}{%
\subsubsection{ARIMA}\label{arima-1}}

The range of parameters that were selected keeping the other constant
are:

\begin{itemize}
\tightlist
\item
  \emph{AR degree (p):} from 0 to 6
\item
  \emph{Integrated degree (d):} 1, as we perform model training on
  closing prices
\item
  \emph{MA degree (p):} from 0 to 6
\end{itemize}

The models are chosen based on the Akaike Information Criterion (AIC).
The objective of AIC is to find a balanced model that does not lose a
lot of information and is also accurate. AIC also penalizes the models
with more beta parameters. Therefore, the model with the lowest AIC is
chosen. Based on Al-Gounmeein and Ismail (2020), AIC is calculated in
the following way, \[
AIC = - 2ln(\hat{l}) + 2k \tag{25} 
\] where: \newline \(k\) - the number of the parameters to be estimated
\newline \(l\) - the likelihood for the respective model

\hypertarget{lstm-1}{%
\subsubsection{LSTM}\label{lstm-1}}

The set of hyperparameters chosen to perform a random search, keeping
the other constant, are the following:

\begin{itemize}
\tightlist
\item
  \emph{Neurons:} {[}25, 50, 75, 100, 250, 500{]}
\item
  \emph{Number of hidden layers:} {[}1, 2{]}
\item
  \emph{Dropout rate:} 0.075
\item
  \emph{Optimizer:} {[}Adam, Nadam, Adagrad{]}
\item
  \emph{Learning rates:} {[}0.01, 0.0001{]}
\item
  \emph{Loss Function:} Mean Square Error
\item
  \emph{Batch size:} 32
\item
  \emph{Sequence Length:} {[}7, 14, 21{]}
\item
  \emph{Input Layer Activation Function:} \(sigmoid\)
\item
  \emph{Output Layer Activation Function:} \(tanh\)
\end{itemize}
The following features were used to predict the closing price at time
\(t+sequence\_length\):

\begin{itemize}
\tightlist
\item
  Closing price at time \(t\)
\item
  Volatility at time \(t\)
\item
  Trading volume at time \(t\)
\end{itemize}

\hypertarget{lstm-arima-1}{%
\subsubsection{LSTM-ARIMA}\label{lstm-arima-1}}

The hyperparameters for the LSTM-ARIMA hybrid model are the same as
those utilized for the individual ARIMA and LSTM models. Random search
serves as the search algorithm to tune these hyperparameters. Initially,
ARIMA generates predictions, which are subsequently incorporated into
the LSTM model. The input layer uses variables such as closing price,
volatility \(t\), trading volume at time \(t\), and residuals from the
ARIMA model at time \(t\) to predict the closing price at time
\(t+sequence\_length\).

\hypertarget{empirical-results}{%
\section{Empirical Results}\label{empirical-results}}

\hypertarget{base-case-results}{%
\subsection{Base case results}\label{base-case-results}}

We evaluate the effectiveness of our investment algorithm using
out-of-sample data. The S\&P 500 equity index began trading on
2005-01-25, the FTSE 100 equity index on 2005-01-13, and the CAC 40
equity index on 2004-12-28, with the trading period continuing until
2023-08-30. Our primary performance evaluation metric is the Modified
Information Ratio (\(IR^{**}\)), as outlined in \emph{Eq. 20}. This
metric offers a comprehensive assessment, encompassing factors such as
annualized return compounded (ARC), return volatility (ASD), and the
largest percentage loss experienced by the asset from its peak value
before reaching a new peak (MD). This approach allows us to not only
assess profitability but also the associated investment risk. For each
equity index, we evaluate three algorithms, namely ARIMA, LSTM, and
LSTM-ARIMA, by comparing them both among themselves and against the
Buy\&Hold strategy.

The intervals of the walk-forward optimization process are uniform for
all equity indices since they all involve daily frequency. The training
period spans 1000 trading days, followed by 250 trading days validation
period and a subsequent 250 trading days testing period. Additionally,
we consider two strategies: \emph{Long-Only}, where only long positions
are allowed, and \emph{Long-Short}, where both long and short positions
are permitted. The research flow is also detailed in section \emph{4.6}
for reference.

\pagebreak

Figure 5, presents the S\&P 500 equity index equity curves of all the
algorithms for both Long-Only and Long-Short strategies respectively and
Table 2 presents the performance metrics for S\&P 500 equity index.
Based on the evaluation of \(IR^{**}\) metrics for both the Long-Only
and Long-Short strategies, it can be deduced that the LSTM-ARIMA
algorithm outperformed the other algorithms. Additionally, it is
noteworthy that all algorithms demonstrated robust performance during
the economic downturn of 2008. Furthermore, a remarkable surge in
performance is observed during the COVID-19 period in the Long-Short
strategy, especially when employing the LSTM-ARIMA algorithm.

\begin{figure}
\centering
\includegraphics[width=0.8\textwidth]{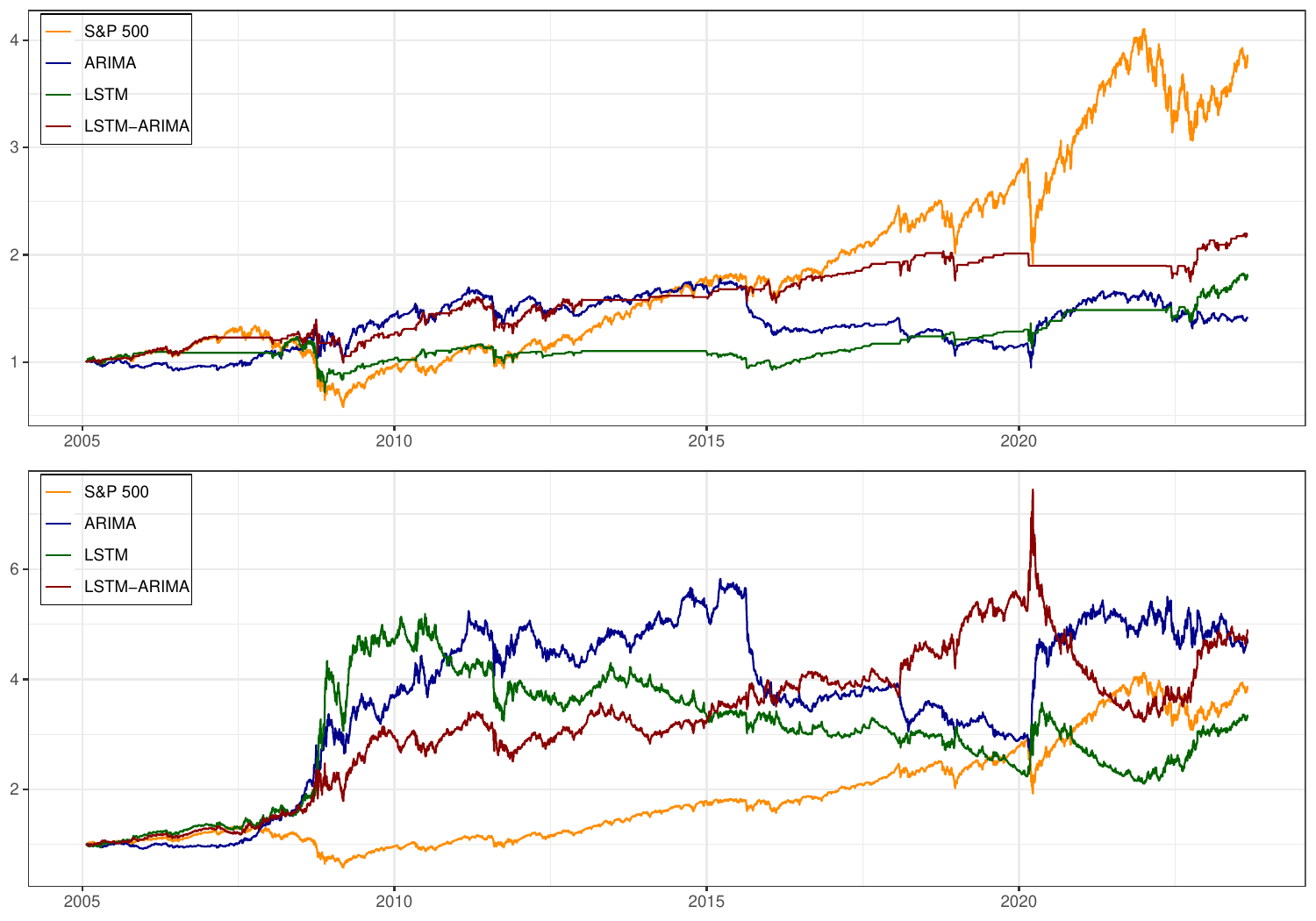}
\caption{The Long-Only and Long-Short Strategy on S\&P 500}
\end{figure}

\vspace{-0.5cm}

\noindent\linespread{0.65}\selectfont {\scriptsize \textbf{Note}: \textit{S\&P 500 is the Buy\&Hold Strategy. The first plot presents the equity curve for the Long-Only strategy and the second plot presents the equity curve for the Long-Short strategy. The trading starts from 2005-01-25. Each equity curve consists of daily frequency data. The transaction costs are 0.1\%.}}

\linespread{1.5}\selectfont

\renewcommand{\arraystretch}{0.75}
\fontsize{8}{8.5}\selectfont

\begin{longtable}[]{@{}llllllll@{}}
\caption{Performance metrics for S\&P 500}\tabularnewline
\toprule\noalign{}
& & ARC(\%) & ASD(\%) & MD(\%) & MLD & IR*(\%) & IR**(\%) \\
\midrule\noalign{}
\endfirsthead
\toprule\noalign{}
& & ARC(\%) & ASD(\%) & MD(\%) & MLD & IR*(\%) & IR**(\%) \\
\midrule\noalign{}
\endhead
\bottomrule\noalign{}
\endlastfoot
\textbf{Long Only} & & & & & & & \\
& \textbf{S\&P 500} & \textbf{7.52} & 19.58 & 56.78 & \textbf{1.65} &
38.43 & 5.09 \\
& ARIMA & 1.89 & 14.45 & 46.73 & 8.45 & 13.07 & 0.53 \\
& LSTM & 3.26 & 13.14 & 41.83 & 9.8 & 24.83 & 1.94 \\
& LSTM-ARIMA & 4.32 & \textbf{11.14} & \textbf{28.95} & 1.67 &
\textbf{38.79} & \textbf{5.79} \\
\textbf{Long Short} & & & & & & & \\
& \textbf{S\&P 500} & 7.52 & 19.58 & 56.78 & \textbf{1.65} & 38.43 &
5.09 \\
& ARIMA & 8.66 & \textbf{19.19} & \textbf{54.81} & 8.44 & 45.11 &
7.13 \\
& LSTM & 6.71 & 19.59 & 59.44 & 13.16 & 34.27 & 3.87 \\
& LSTM-ARIMA & \textbf{8.92} & 19.58 & 56.62 & 3.44 & \textbf{45.56} &
\textbf{7.18} \\
\end{longtable}

\normalsize
\vspace{-0.3cm}

\noindent\linespread{0.65}\selectfont {\scriptsize \textbf{Note}: \textit{S\&P 500 represents the benchmark Buy\&Hold Strategy. Trading starts from 2005-01-25. The transaction costs are 0.1\%. The best strategy is the one that holds the highest Modified Information Ratio ($IR^{**}$). Columns with the best corresponding values are denoted in bold.}}

\linespread{1.5}\selectfont

\pagebreak

Figure 6 presents the equity curve of FTSE 100 equity index and Table 3
presents the performance metrics for FTSE 100 equity index. Based on the
evaluation of \(IR^{**}\) metrics for both the Long-Only and Long-Short
strategies, it can be deduced that the LSTM-ARIMA algorithm outperformed
the other algorithms for the FTSE 100 equity index. Both the Long-Only
and Long-Short strategies, when executed with the LSTM-ARIMA algorithm,
display a substantial peak in performance after the year 2020.

\begin{figure}
\centering
\includegraphics[width=0.8\textwidth]{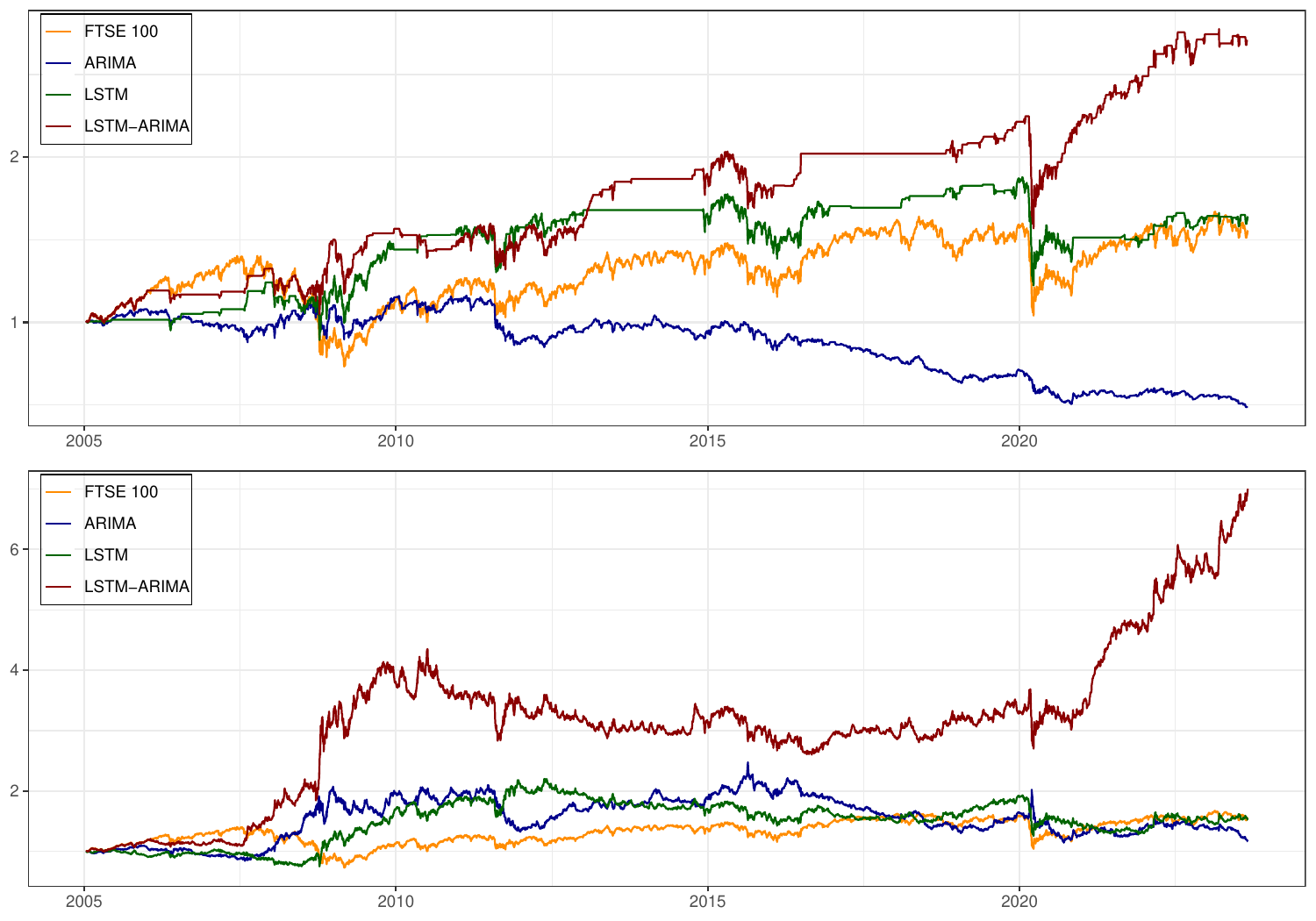}
\caption{The Long-Only and Long-Short Strategy on FTSE 100}
\end{figure}

\vspace{-0.5cm}

\noindent\linespread{0.65}\selectfont {\scriptsize \textbf{Note}: \textit{FTSE 100 represents the benchmark Buy\&Hold Strategy. The first plot presents the equity curve for the Long-Only strategy and the second plot presents the equity curve for the Long-Short strategy. The trading starts from 2005-01-13. Each equity curve consists of daily frequency data. The transaction costs are 0.1\%.}}

\linespread{1.5}\selectfont

\renewcommand{\arraystretch}{0.75}
\fontsize{8}{8.5}\selectfont

\begin{longtable}[]{@{}llcccccc@{}}
\caption{Performance metrics for FTSE 100}\tabularnewline
\toprule\noalign{}
& & ARC(\%) & ASD(\%) & MD(\%) & MLD & IR*(\%) & IR**(\%) \\
\midrule\noalign{}
\endfirsthead
\toprule\noalign{}
& & ARC(\%) & ASD(\%) & MD(\%) & MLD & IR*(\%) & IR**(\%) \\
\midrule\noalign{}
\endhead
\bottomrule\noalign{}
\endlastfoot
\textbf{Long Only} & & & & & & & \\
& \textbf{FTSE 100} & 2.39 & 18.03 & 47.83 & 5.94 & 13.27 & 0.66 \\
& ARIMA & -3.78 & \textbf{12.88} & 58.12 & 12.55 & -29.38 & -1.91 \\
& LSTM & 2.68 & 14.32 & 34.93 & 3.61 & 18.75 & 1.44 \\
& LSTM-ARIMA & \textbf{5.47} & 13.79 & \textbf{30.22} & \textbf{0.91} &
\textbf{39.71} & \textbf{7.19} \\
\textbf{Long Short} & & & & & & & \\
& \textbf{FTSE 100} & 2.39 & 18.03 & 47.83 & \textbf{5.94} & 13.27 &
0.66 \\
& ARIMA & 0.84 & 18.04 & 53.65 & 8.03 & 4.66 & 0.07 \\
& LSTM & 2.28 & 18.03 & 42.92 & 11.3 & 12.67 & 0.67 \\
& LSTM-ARIMA & \textbf{10.98} & \textbf{18.02} & \textbf{40.17} & 10.89
& \textbf{60.92} & \textbf{16.65} \\
\end{longtable}

\normalsize
\vspace{-0.3cm}

\noindent\linespread{0.65}\selectfont {\scriptsize \textbf{Note}: \textit{FTSE 100 represents the benchmark Buy\&Hold Strategy. Trading starts from 2005-01-13. The transaction costs are 0.1\%. The best strategy is the one that holds the highest Modified Information Ratio ($IR^{**}$). Columns with the best corresponding values are denoted in bold.}}

\linespread{1.5}\selectfont

\pagebreak

Figure 7 presents the equity curve of the CAC 40 equity index and Table
4 presents the performance metrics for CAC 40 equity index. Based on the
evaluation of \(IR^{**}\) metrics for both the Long-Only and Long-Short
strategies, it can be deduced that the LSTM-ARIMA algorithm outperformed
the other algorithms for the CAC 40 equity index. Furthermore, we
noticed that both Long-Only and Long-Short strategies, when executed
with the LSTM-ARIMA algorithm, achieved a high return during the
post-Covid time.

\begin{figure}
\centering
\includegraphics[width=0.8\textwidth]{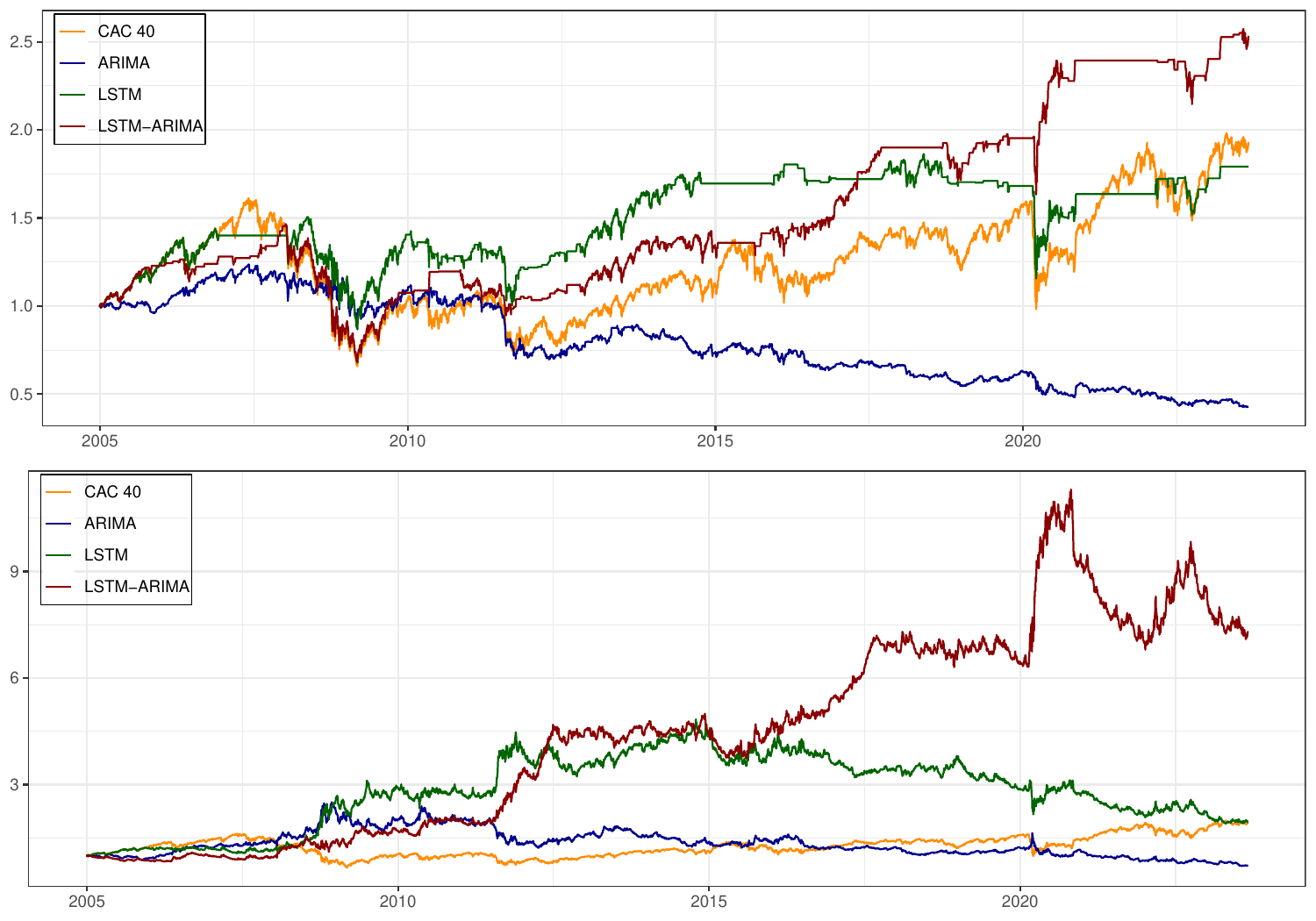}
\caption{The Long-Only and Long-Short Strategy on CAC 40}
\end{figure}

\vspace{-0.5cm}

\noindent\linespread{0.65}\selectfont {\scriptsize \textbf{Note}: \textit{CAC 40 represents the benchmark Buy\&Hold Strategy. The first plot presents the equity curve for the Long-Only strategy and the second plot presents the equity curve for the Long-Short strategy. The trading starts from 2004-12-28. Each equity curve consists of daily frequency data. The transaction costs are 0.1\%.}}

\linespread{1.5}\selectfont

\renewcommand{\arraystretch}{0.75}
\fontsize{8}{8.5}\selectfont

\begin{longtable}[]{@{}llcccccc@{}}
\caption{Performance statistics for CAC 40}\tabularnewline
\toprule\noalign{}
& & ARC(\%) & ASD(\%) & MD(\%) & MLD & IR*(\%) & IR**(\%) \\
\midrule\noalign{}
\endfirsthead
\toprule\noalign{}
& & ARC(\%) & ASD(\%) & MD(\%) & MLD & IR*(\%) & IR**(\%) \\
\midrule\noalign{}
\endhead
\bottomrule\noalign{}
\endlastfoot
\textbf{Long Only} & & & & & & & \\
& \textbf{CAC 40} & 3.52 & 21.44 & 59.16 & 14.04 & 16.43 & 0.98 \\
& ARIMA & -4.38 & \textbf{15.14} & 65.53 & 16.5 & -28.9 & -1.93 \\
& LSTM & 3.12 & 16.1 & \textbf{42.35} & \textbf{5.38} & 19.4 & 1.43 \\
& LSTM-ARIMA & \textbf{5.02} & 15.43 & 53.65 & 8.33 & \textbf{32.52} &
\textbf{3.04} \\
\textbf{Long Short} & & & & & & & \\
& \textbf{CAC 40} & 3.52 & 21.44 & 59.16 & 14.04 & 16.43 & 0.98 \\
& ARIMA & -1.81 & \textbf{21.43} & 72.02 & 14.95 & -8.45 & -0.21 \\
& LSTM & 3.56 & 21.44 & 60.73 & 9.01 & 16.59 & 0.97 \\
& LSTM-ARIMA & \textbf{11.06} & \textbf{21.43} & \textbf{39.91} &
\textbf{2.91} & \textbf{51.6} & \textbf{14.29} \\
\end{longtable}

\normalsize
\vspace{-0.3cm}

\noindent\linespread{0.65}\selectfont {\scriptsize \textbf{Note}: \textit{CAC 40 represents the benchmark Buy\&Hold Strategy. Trading starts from 2004-12-28. The transaction costs are 0.1\%. The best strategy is the one that holds the highest Modified Information Ratio ($IR^{**}$). Columns with the best corresponding values are denoted in bold.}}

\linespread{1.5}\selectfont

\pagebreak

\hypertarget{statistical-significance}{%
\subsection{Statistical Significance}\label{statistical-significance}}

While LSTM-ARIMA outperformed all the other algorithms for both
Long-Only and Long-Short strategies, it would be premature to conclude
that the expected values of these strategies' returns distributions
surpass those of the benchmarks. Hence, it is prudent to subject them to
statistical inference testing to validate their efficacy. We test it
using a t-test for paired samples (Devore and Berk (2012)) with the
following hypotheses: \[ \begin{cases}  
H_0: \mu_{d} = \mu_{strategy} - \mu_{benchmark} = 0 \\
H_1: \mu_{d} > 0
\end{cases}
\tag{26} \]
where: 
\newline \(\mu_{strategy}\) - the expected value of the
strategy \newline \(\mu_{benchmark}\) - the expected value of the
benchmark \newline \(\mu_d\) - the difference between the expected
values of the strategy and benchmark returns \newline

\renewcommand{\arraystretch}{0.75}
\fontsize{8}{8.5}\selectfont

\begin{longtable}[]{@{}llllll@{}}
\caption{P-values for the paired t-test}\tabularnewline
\toprule\noalign{}
& & & \textbf{S\&P 500} & \textbf{FTSE 100} & \textbf{CAC 40} \\
\midrule\noalign{}
\endfirsthead
\toprule\noalign{}
& & & \textbf{S\&P 500} & \textbf{FTSE 100} & \textbf{CAC 40} \\
\midrule\noalign{}
\endhead
\bottomrule\noalign{}
\endlastfoot
\textbf{Long-Only} & & & & & \\
& & ARIMA & \textbf{0.0362} & \textbf{0.0152} & \textbf{0.0086} \\
& & LSTM & 0.1248 & 0.8964 & 0.6696 \\
& & LSTM-ARIMA & 0.2420 & 0.3911 & 0.9249 \\
\textbf{Long-Short} & & & & & \\
& & ARIMA & 0.8730 & 0.7902 & 0.4455 \\
& & LSTM & 0.9046 & 0.9784 & 0.9961 \\
& & LSTM-ARIMA & 0.8569 & 0.1709 & 0.3310 \\
\end{longtable}

\normalsize
\vspace{-0.3cm}

\noindent\linespread{0.65}\selectfont {\scriptsize \textbf{Note}: \textit{The table presents the p-values of the paired t-test. The significance level is set at 10\%. P-values less than 0.1 are in bold. Each strategy has been compared with the benchmark Buy\&Hold strategy. S\&P 500, FTSE 100, and CAC 40 represent the benchmark Buy\&Hold Strategy.}}

\linespread{1.5}\selectfont

Our significance level is set at 10\%. If the p-value is lower than 0.1,
we reject the null hypotheses; otherwise, we have no grounds to reject
it. Based on the p-values provided in Table 5, we infer that the only
statistically significant results we attain are observed for the ARIMA
model with the Long-Only strategy for all equity indices. In all other
scenarios, our findings lack statistical significance. Nonetheless, we
opt to conduct an additional test. We formulate a linear regression
model as described in eq. 27 and subsequently carry out a right-sided
t-test (Wooldridge (2015)) to assess the significance of the intercept,
as shown in eq. 28.

\pagebreak

\[ R_{strategy} = \alpha + \beta \space\times\space r_{benchmark} + \epsilon_{t} \tag{27}\]
\[ \begin{cases}  
H_0: \alpha = 0 \\
H_1: \alpha > 0
\end{cases}
\tag{28} \]

\renewcommand{\arraystretch}{0.75}
\fontsize{9}{9.5}\selectfont

\fontsize{7}{7.5}\selectfont
\begin{longtable}{@{}lllllllllll@{}}
\caption{Simple Linear Regression results} \\
\toprule
& & & α & SE(α) & $t_\alpha$ & $p_\alpha$ & β & SE(β) & $t_\beta$ & $p_\beta$ \\
\midrule
\endfirsthead
\toprule
& & & α & SE(α) & $t_\alpha$ & $p_\alpha$ & β & SE(β) & $t_\beta$ & $p_\beta$ \\
\midrule
\endhead
\bottomrule
\endlastfoot
\textbf{Long-Only} & & & & & & & & & & \\
& \textbf{S\&P 500} & & & & & & & & & \\
& & ARIMA & -0.0001 & 0.0001 & -0.9864 & 0.8380 & 0.5547 & 0.0071 & 77.9073 & 0.0000 \\
& & LSTM & 0.0000 & 0.0001 & -0.0504 & 0.5201 & 0.4554 & 0.0072 & 63.1896 & 0.0000 \\
& & LSTM-ARIMA & 0.0001 & 0.0001 & 0.8697 & 0.1923 & 0.3274 & 0.0068 & 48.1206 & 0.0000 \\
& \textbf{FTSE 100} & & & & & & & & & \\
& & ARIMA & -0.0002 & 0.0001 & -2.4608 & 0.9930 & 0.5146 & 0.0072 & 71.2718 & 0.0000 \\
& & LSTM & 0.0000 & 0.0001 & 0.5702 & 0.2843 & 0.6344 & 0.0070 & 91.1631 & 0.0000 \\
& & LSTM-ARIMA & 0.0002 & 0.0001 & 1.9299 & \textbf{0.0268} & 0.5894 & 0.0071 & 83.0157 & 0.0000 \\
& \textbf{CAC 40} & & & & & & & & & \\
& & ARIMA & -0.0002 & 0.0001 & -2.5518 & 0.9946 & 0.5033 & 0.0072 & 70.1771 & 0.0000 \\
& & LSTM & 0.0000 & 0.0001 & 0.4595 & 0.3230 & 0.5647 & 0.0072 & 78.8092 & 0.0000 \\
& & LSTM-ARIMA & 0.0001 & 0.0001 & 1.2578 & 0.1043 & 0.5189 & 0.0072 & 71.8728 & 0.0000 \\
\textbf{Long-Short} & & & & & & & & & & \\
& \textbf{S\&P 500} & & & & & & & & & \\
& & ARIMA & 0.0004 & 0.0002 & 2.0281 & \textbf{0.0213} & 0.1292 & 0.0142 & 9.0957 & 0.0000 \\
& & LSTM & 0.0004 & 0.0002 & 2.0726 & \textbf{0.0191} & -0.1069 & 0.0145 & -7.3498 & 1.0000 \\
& & LSTM-ARIMA & 0.0005 & 0.0002 & 2.6172 & \textbf{0.0044} & -0.1448 & 0.0145 & -10.0100 & 1.0000 \\
& \textbf{FTSE 100} & & & & & & & & & \\
& & ARIMA & 0.0001 & 0.0002 & 0.5591 & 0.2881 & 0.0301 & 0.0146 & 2.0670 & 0.0194 \\
& & LSTM & 0.0001 & 0.0002 & 0.6003 & 0.2742 & 0.3932 & 0.0134 & 29.3379 & 0.0000 \\
& & LSTM-ARIMA & 0.0005 & 0.0002 & 2.8736 & \textbf{0.0020} & 0.0115 & 0.0146 & 0.7874 & 0.2155 \\
& \textbf{CAC 40} & & & & & & & & & \\
& & ARIMA & 0.0000 & 0.0002 & 0.0855 & 0.4659 & 0.0079 & 0.0145 & 0.5432 & 0.2935 \\
& & LSTM & 0.0003 & 0.0002 & 1.3807 & \textbf{0.0837} & -0.1596 & 0.0143 & -11.1765 & 1.0000 \\
& & LSTM-ARIMA & 0.0005 & 0.0002 & 2.6859 & \textbf{0.0036} & -0.0728 & 0.0144 & -5.0480 & 1.0000 \\
\end{longtable}
\fontsize{9}{9.5}\selectfont
\normalsize
\vspace{-0.3cm}

\noindent\linespread{0.65}\selectfont {\scriptsize \textbf{Note}: \textit{The table presents the results of the linear regression as mentioned in eq. 28. The significance level is set at 10\%. P-values less than 0.1 are in bold. Each strategy has been compared with the benchmark Buy\&Hold strategy. S\&P 500, FTSE 100, and CAC 40 represent the benchmark Buy\&Hold Strategy.}}

\linespread{1.5}\selectfont

The findings from Table 6 details the outcomes of the simple linear
regression analysis we conducted.\} Based on \(p_\alpha\), statistically
significant algorithms were observed at a significance level of 10\%.
Specifically, for the Long-Only strategy, LSTM-ARIMA for the FTSE 100
equity index exhibited statistical significance. Regarding the
Long-Short strategy, all algorithms demonstrated statistical
significance for the S\&P 500 equity index. For the FTSE 100 equity
index, only LSTM-ARIMA demonstrated statistical significance. Finally,
in the case of the CAC 40 equity index, both LSTM and LSTM-ARIMA
exhibited statistical significance.

\pagebreak

\hypertarget{summary}{%
\subsection{Summary}\label{summary}}

Based on the results presented in Table 2, we see that our novel
LSTM-ARIMA algorithm outperformed all the other algorithms for both
\textbf{Long-Only} and \textbf{Long-Short} strategies and all the equity
indices. In the case of the S\&P 500 equity index, the \textbf{Long-Only
strategy for LSTM-ARIMA algorithm} obtained a modified information ratio
(\(IR^{**}\)) of \emph{5.79\%} and the \textbf{Long-Short strategy for
LSTM-ARIMA algorithm} of \emph{7.18\%}. In the case of the FTSE 100
equity index, the \textbf{Long-Only strategy for LSTM-ARIMA algorithm}
obtained an \(IR^{**}\) of \emph{7.19\%} and the \textbf{Long-Short
strategy for LSTM-ARIMA algorithm} of \emph{16.65\%}. Finally, in the
case of the CAC 40 equity index, the \textbf{Long-Only strategy for
LSTM-ARIMA algorithm} obtained an \(IR^{**}\) of \emph{3.04\%} and the
\textbf{Long-Short strategy for LSTM-ARIMA algorithm} of \emph{14.29\%}.

Furthermore, based on the results from the previous section, the paired
t-test (Table 5) showed that only statistically significant results were
observed for the ARIMA model with the Long-Only strategy for all the
equity indices. However, summarizing the outcomes of the simple linear
regression analysis (Table 6), it becomes evident that the intercept of
the LSTM-ARIMA algorithm with the Long-Short strategy returns,
significantly exceeded 0 when regressed against the returns from the
Buy\&Hold strategy.

\pagebreak

\hypertarget{sensitivity-analysis}{%
\section{Sensitivity Analysis}\label{sensitivity-analysis}}

This section is specifically focused on addressing the third research
question (\emph{RQ3}). Its primary objective is to investigate how
changes in specific parameters and hyperparameters influence the output.
Through this assessment, we can evaluate the stability and reliability
of our investment algorithm. In the case of the ARIMA model, we alter
the following parameters:

\begin{itemize}
\tightlist
\item
  The range of ARIMA model order: (p, d, q) = (0-3, 1, 0-3)
\item
  The information criterion: Bayesian Information Criterion (BIC)
\end{itemize}
The following parameters are altered for LSTM and LSTM-ARIMA:

\begin{itemize}
\tightlist
\item
  \emph{the Dropout Rate:} 0.05, 0.1
\item
  \emph{the batch size:} 16, 64
\end{itemize}

During the sensitivity analysis, only the parameters mentioned are
changed and everything else is kept as they were. We continue to use the
modified information ratio (\(IR^{**}\)) as the main evaluation metric.
Additionally, this section is divided into three subsections by our
algorithms.

\hypertarget{arima-2}{%
\subsection{ARIMA}\label{arima-2}}

Figure 8 and Table 7 present the sensitivity analysis results for S\&P
500, FTSE 100, and CAC 40 equity indices. In the case of the S\&P 500
equity index, based on the \(IR^{**}\) metrics, the base case
outperforms all the other changes in parameters for the Long-Only
strategy. However, under the Long-Short strategy, we notice enhancements
in the results when employing the Bayesian Information Criterion (BIC)
as the information criterion. In the case of the FTSE 100 equity index,
the performance of the ARIMA model appears to be poor. Albeit, for the
Long-Short strategy, improvements were seen when narrowing the range of
the orders. The model's performance benefited from this adjustment,
possibly suggesting that high-order configurations may have been
predisposed to overfitting. In the case of the CAC 40 equity index, the
changes in the parameters do not yield improved results. The Buy\&Hold
still outperforms the ARIMA model.

In summary, the results of the ARIMA model used in our algorithmic
investment strategy exhibit robustness to changes in the information
criterion and the order range setting of the model.

\pagebreak

\begin{figure}
\centering
\includegraphics[width=1\textwidth]{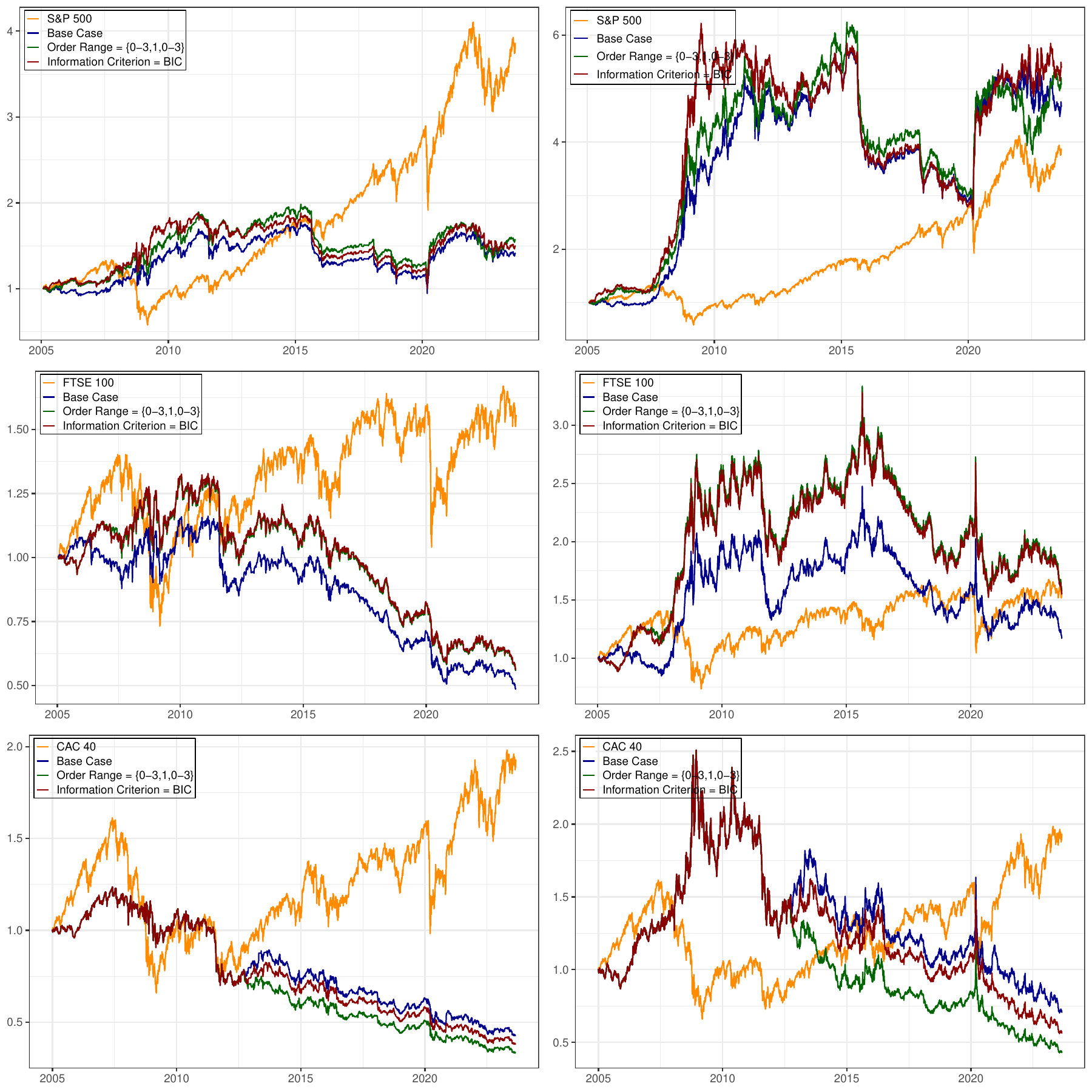}
\caption{ARIMA Sensitivity Analysis}
\end{figure}

\vspace{-0.3cm}

\noindent\linespread{0.65}\selectfont {\scriptsize \textbf{Note}: \textit{The figure presents the equity curves for the sensitivity analysis performed on the ARIMA model. S\&P 500, FTSE 100, and CAC 40 represent the benchmark Buy\&Hold strategy for each index respectively. The base case scenario utilizes the order range (p,d,q)={0-6, 1, 0-6} and akaike information criterion (AIC). S\&P 500 index trading starts on 2005-01-25, FTSE 100 equity index trading starts on 2005-01-13, and CAC 40 equity index trading starts on 2004-12-28. Each equity curve consists of daily frequency data. The transaction costs are 0.1\%. The best values are in bold and are bolded with respect to the base case scenario.}}

\linespread{1.5}\selectfont

\pagebreak

\renewcommand{\arraystretch}{0.75}
\fontsize{10}{10.5}\selectfont

\begin{longtable}[]{@{}llcccccc@{}}
\caption{ARIMA Sensitivity Analysis performance metrics}\tabularnewline
\toprule\noalign{}
& & ARC(\%) & ASD(\%) & MD(\%) & MLD & IR*(\%) & IR**(\%) \\
\midrule\noalign{}
\endfirsthead
\toprule\noalign{}
& & ARC(\%) & ASD(\%) & MD(\%) & MLD & IR*(\%) & IR**(\%) \\
\midrule\noalign{}
\endhead
\bottomrule\noalign{}
\endlastfoot
& \textbf{S\&P 500} & 7.52 & 19.58 & 56.78 & 1.65 & 38.43 & 5.09 \\
& ~ & & & & & & \\
\textbf{Long Only} & Base Case & \textbf{4.32} & \textbf{11.14} &
\textbf{28.95} & \textbf{1.67} & \textbf{38.79} & \textbf{5.79} \\
& Order Range = \{0-3,1,0-3\} & 2.47 & 14.45 & 46.73 & 8.45 & 17.11 &
0.91 \\
& Information Criterion = BIC & 2.21 & 14.35 & 46.73 & 8.45 & 15.41 &
0.73 \\
& ~ & & & & & & \\
\textbf{Long Short} & Base Case & 8.92 & 19.58 & 56.62 & \textbf{3.44} &
45.56 & 7.18 \\
& Order Range = \{0-3,1,0-3\} & 9.15 & \textbf{19.57} & \textbf{56.24} &
8.44 & 46.79 & 7.62 \\
& Information Criterion = BIC & \textbf{9.52} & \textbf{19.57} & 58.85 &
14.18 & \textbf{48.62} & \textbf{7.86} \\
& ~ & & & & & & \\
& \textbf{FTSE 100} & 2.39 & 18.03 & 47.83 & 5.94 & 13.27 & 0.66 \\
& ~ & & & & & & \\
\textbf{Long Only} & Base Case & -3.78 & \textbf{12.88} & 58.12 &
\textbf{12.55} & -29.38 & -1.91 \\
& Order Range = \{0-3,1,0-3\} & -3.06 & 12.9 & \textbf{57.55} &
\textbf{12.55} & -23.71 & -1.26 \\
& Information Criterion = BIC & \textbf{-3.02} & 12.9 & \textbf{57.55} &
\textbf{12.55} & \textbf{-23.4} & \textbf{-1.23} \\
& ~ & & & & & & \\
\textbf{Long Short} & Base Case & 0.84 & \textbf{18.04} & \textbf{53.65}
& \textbf{8.03} & 4.66 & 0.07 \\
& Order Range = \{0-3,1,0-3\} & \textbf{2.46} & \textbf{18.04} &
\textbf{53.65} & \textbf{8.03} & \textbf{13.66} & \textbf{0.63} \\
& Information Criterion = BIC & 2.37 & \textbf{18.04} & \textbf{53.65} &
\textbf{8.03} & 13.15 & 0.58 \\
& ~ & & & & & & \\
& \textbf{CAC 40} & 3.52 & 21.44 & 59.16 & 14.04 & 16.43 & 0.98 \\
& ~ & & & & & & \\
\textbf{Long Only} & Base Case & \textbf{-4.38} & 15.14 & \textbf{65.53}
& \textbf{16.5} & \textbf{-28.9} & \textbf{-1.93} \\
& Order Range = \{0-3,1,0-3\} & -5.62 & 15.1 & 73.12 & \textbf{16.5} &
-37.24 & -2.86 \\
& Information Criterion = BIC & -4.95 & \textbf{15.07} & 69.27 &
\textbf{16.5} & -32.86 & -2.35 \\
& ~ & & & & & & \\
\textbf{Long Short} & Base Case & \textbf{-1.81} & \textbf{21.43} &
\textbf{72.02} & \textbf{14.95} & \textbf{-8.45} & \textbf{-0.21} \\
& Order Range = \{0-3,1,0-3\} & -4.35 & \textbf{21.43} & 82.97 &
\textbf{14.95} & -20.28 & -1.06 \\
& Information Criterion = BIC & -2.97 & \textbf{21.43} & 77.67 &
\textbf{14.95} & -13.86 & -0.53 \\
& ~ & & & & & & \\
\end{longtable}

\normalsize
\vspace{-0.3cm}

\noindent\linespread{0.65}\selectfont {\scriptsize \textbf{Note}: \textit{The table shows the performance metrics for the sensitivity analysis performed on the ARIMA model. S\&P 500, FTSE 100, and CAC 40 represent the benchmark Buy\&Hold strategy for each index respectively. S\&P 500 index trading starts on 2005-01-25, FTSE 100 equity index trading starts on 2005-01-13, and CAC 40 equity index trading starts on 2004-12-28. In the base case scenario, the Dropout rate is set to 0.075 and the Batch Size is set to 32. The transaction costs are 0.1\%. The best values are in bold and are bolded with respect to the base case scenario.}}

\linespread{1.5}\selectfont

\pagebreak

\hypertarget{lstm-2}{%
\subsection{LSTM}\label{lstm-2}}

Figure 9 and Table 8 present the results of a sensitivity analysis
conducted on the S\&P 500, FTSE 100, and CAC 40 equity indices using the
LSTM model. In the case of the S\&P 500 equity index (presented in Panel
A and Panel C), it is observed that an increase in the dropout rate
results in a higher modified information ratio (\(IR^{**}\)) for both
the Long-Only and the Long-Short strategy. Additionally, as indicated in
Panel B and Panel D, it is inferred that a smaller batch size yields a
higher \(IR^{**}\) for the Long-Only strategy, whereas for the
Long-Short strategy smaller batch size made it significantly worse.

In the case of the FTSE 100 equity index, indicated in Panel E and Panel
G, it is evident that smaller dropout rates yield better results in the
case of the Long-Only strategy, while higher dropout rates are more
effective for the Long-Short strategy. Furthermore, as indicated in
Panel F and Panel H, it is apparent that the Long-Short strategy
exhibits higher \(IR^{**}\) when using a smaller batch size, while the
Long-Only strategy achieves optimal performance when utilizing the base
case scenario.

The results for the CAC 40 equity index, presented in Panel I and Panel
K, indicate that a smaller dropout rate leads to a higher \(IR^{**}\)
for the Long-Only strategy. However, this trend is not observed for the
Long-Short strategy, as the base case scenario continues to deliver the
best results. Furthermore, based on the observations in Panel J and
Panel L, it is noticed that a higher batch size results in higher
annualized returns compounded (\(ARC\)) for both the Long-Only and the
Long-Short strategies. However, when considering the \(IR^{**}\) metric,
the improvement is only evident in the Long-Short strategy, whereas the
Long-Only strategy achieves a higher \(IR^{**}\) with a smaller batch
size.

\pagebreak

\begin{figure}
\centering
\includegraphics{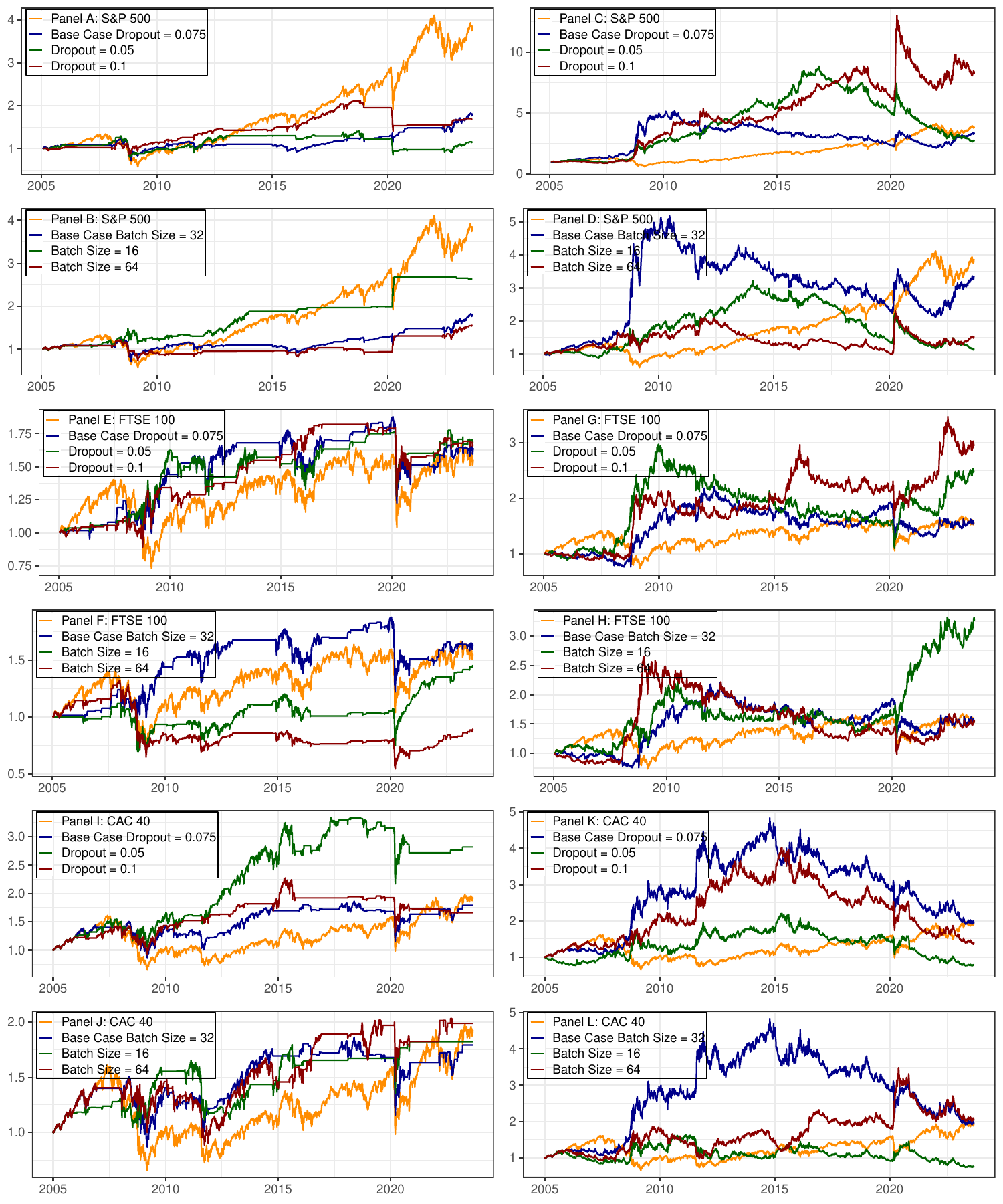}
\caption{LSTM Sensitivity Analysis for S\&P 500}
\end{figure}

\vspace{-0.3cm}

\noindent\linespread{0.65}\selectfont {\scriptsize \textbf{Note}: \textit{The figure presents the equity curves for the sensitivity analysis performed on the LSTM model. S\&P 500, FTSE 100, and CAC 40 represent the benchmark Buy\&Hold strategy for each index respectively. S\&P 500 index trading starts on 2005-01-25, FTSE 100 equity index trading starts on 2005-01-13, and CAC 40 equity index trading starts on 2004-12-28. Each equity curve consists of daily frequency data. The transaction costs are 0.1\%. The best values are in bold and are bolded with respect to the base case scenario.}}

\linespread{1.5}\selectfont

\pagebreak

\renewcommand{\arraystretch}{0.72}
\fontsize{9}{9.5}\selectfont

\begin{longtable}[]{@{}llcccccc@{}}
\caption{LSTM Sensitivity Analysis performance metrics}\tabularnewline
\toprule\noalign{}
& & ARC(\%) & ASD(\%) & MD(\%) & MLD & IR*(\%) & IR**(\%) \\
\midrule\noalign{}
\endfirsthead
\toprule\noalign{}
& & ARC(\%) & ASD(\%) & MD(\%) & MLD & IR*(\%) & IR**(\%) \\
\midrule\noalign{}
\endhead
\bottomrule\noalign{}
\endlastfoot
~ & \textbf{S\&P 500} & 7.52 & 19.58 & 56.78 & 1.65 & 38.43 & 5.09 \\
~ & ~ & & & & & & \\
\textbf{Long Only} & Base Case (Dropout = 0.075) & \textbf{3.26} & 13.14
& 41.83 & 9.8 & 24.83 & 1.94 \\
\textbf{Panel A: Dropout Rate} & Dropout = 0.05 & 0.79 & 12.68 & 39.43 &
4.94 & 6.24 & 0.13 \\
& Dropout = 0.1 & 2.87 & \textbf{11.4} & \textbf{33.93} & \textbf{4.86}
& \textbf{25.2} & \textbf{2.13} \\
& ~ & & & & & & \\
\textbf{Panel B: Batch Size} & Base Case (Batch Size = 32) & 3.26 &
13.14 & 41.83 & 9.8 & 24.83 & 1.94 \\
& Batch Size = 16 & \textbf{5.37} & \textbf{10.45} & \textbf{24.07} &
\textbf{3.96} & \textbf{51.37} & \textbf{11.46} \\
& Batch Size = 64 & 2.42 & 10.94 & 38.72 & 11.77 & 22.14 & 1.39 \\
& ~ & & & & & & \\
\textbf{Long Short} & Base Case (Dropout = 0.075) & 6.71 &
\textbf{19.59} & 59.44 & 13.16 & 34.27 & 3.87 \\
\textbf{Panel C: Dropout Rate} & Dropout = 0.05 & 5.42 & \textbf{19.59}
& 70.44 & 6.8 & 27.66 & 2.13 \\
& Dropout = 0.1 & \textbf{12.0} & \textbf{19.59} & \textbf{47.23} &
\textbf{3.4} & \textbf{61.26} & \textbf{15.57} \\
& ~ & & & & & & \\
\textbf{Panel D: Batch Size} & Base Case (Batch Size = 32) &
\textbf{6.71} & 19.59 & 59.44 & 13.16 & \textbf{34.27} &
\textbf{3.87} \\
& Batch Size = 16 & 0.72 & 19.59 & 65.51 & 9.57 & 3.67 & 0.04 \\
& Batch Size = 64 & 2.26 & \textbf{19.58} & \textbf{53.48} &
\textbf{8.23} & 11.55 & 0.49 \\
& ~ & & & & & & \\
~ & \textbf{FTSE 100} & 2.39 & 18.03 & 47.83 & 5.94 & 13.27 & 0.66 \\
~ & ~ & & & & & & \\
\textbf{Long Only} & Base Case (Dropout = 0.075) & 2.68 & 14.32 & 34.93
& 3.61 & 18.75 & 1.44 \\
\textbf{Panel E: Dropout Rate} & Dropout = 0.05 & \textbf{2.85} & 13.1 &
\textbf{27.53} & \textbf{2.45} & 21.78 & \textbf{2.26} \\
& Dropout = 0.1 & 2.78 & \textbf{11.99} & 30.44 & 4.77 & \textbf{23.17}
& 2.12 \\
& ~ & & & & & & \\
\textbf{Panel F: Batch Size} & Base Case (Batch Size = 32) &
\textbf{2.68} & 14.32 & \textbf{34.93} & \textbf{3.61} & \textbf{18.75}
& \textbf{1.44} \\
& Batch Size = 16 & 2.03 & \textbf{13.84} & 37.95 & 5.86 & 14.66 &
0.78 \\
& Batch Size = 64 & -0.59 & 14.09 & 58.72 & 15.68 & -4.2 & -0.04 \\
& ~ & & & & & & \\
\textbf{Long Short} & Base Case (Dropout = 0.075) & 2.28 &
\textbf{18.03} & \textbf{42.92} & 11.3 & 12.67 & 0.67 \\
\textbf{Panel G: Dropout Rate} & Dropout = 0.05 & 4.96 & \textbf{18.03}
& 63.39 & 13.7 & 27.49 & 2.15 \\
& Dropout = 0.1 & \textbf{5.96} & 18.05 & 44.81 & \textbf{6.09} &
\textbf{33.0} & \textbf{4.39} \\
& ~ & & & & & & \\
\textbf{Panel H: Batch Size} & Base Case (Batch Size = 32) & 2.28 &
\textbf{18.03} & \textbf{42.92} & 11.3 & 12.67 & 0.67 \\
& Batch Size = 16 & \textbf{6.53} & \textbf{18.03} & 43.77 &
\textbf{10.2} & \textbf{36.23} & \textbf{5.4} \\
& Batch Size = 64 & 2.28 & 18.04 & 63.08 & 14.69 & 12.63 & 0.46 \\
& ~ & & & & & & \\
~ & \textbf{CAC 40} & 3.52 & 21.44 & 59.16 & 14.04 & 16.43 & 0.98 \\
~ & ~ & & & & & & \\
\textbf{Long Only} & Base Case (Dropout = 0.075) & 3.12 & 16.1 & 42.35 &
\textbf{5.38} & 19.4 & 1.43 \\
\textbf{Panel I: Dropout Rate} & Dropout = 0.05 & \textbf{5.62} & 16.79
& \textbf{34.96} & 5.52 & \textbf{33.48} & \textbf{5.38} \\
& Dropout = 0.1 & 2.71 & \textbf{14.56} & 42.18 & 8.49 & 18.62 & 1.2 \\
& ~ & & & & & & \\
\textbf{Panel J: Batch Size} & Base Case (Batch Size = 32) & 3.12 & 16.1
& 42.35 & 5.38 & 19.4 & 1.43 \\
& Batch Size = 16 & 3.22 & \textbf{13.91} & \textbf{33.09} & 4.06 &
\textbf{23.13} & \textbf{2.25} \\
& Batch Size = 64 & \textbf{3.69} & 16.98 & 40.42 & \textbf{3.6} & 21.74
& 1.99 \\
& ~ & & & & & & \\
\textbf{Long Short} & Base Case (Dropout = 0.075) & \textbf{3.56} &
21.44 & \textbf{60.73} & 9.01 & \textbf{16.59} & \textbf{0.97} \\
\textbf{Panel K: Dropout Rate} & Dropout = 0.05 & -1.34 & \textbf{21.43}
& 65.66 & \textbf{8.49} & -6.26 & -0.13 \\
& Dropout = 0.1 & 1.75 & 21.44 & 66.37 & \textbf{8.49} & 8.16 & 0.22 \\
& ~ & & & & & & \\
\textbf{Panel L: Batch Size} & Base Case (Batch Size = 32) & 3.56 &
21.44 & 60.73 & 9.01 & 16.59 & 0.97 \\
& Batch Size = 16 & -1.5 & 21.46 & 55.1 & 12.72 & -7.0 & -0.19 \\
& Batch Size = 64 & \textbf{3.82} & \textbf{21.43} & \textbf{47.78} &
\textbf{3.35} & \textbf{17.85} & \textbf{1.43} \\
& ~ & & & & & & \\
\end{longtable}

\normalsize
\vspace{-0.3cm}

\noindent\linespread{0.65}\selectfont {\scriptsize \textbf{Note}: \textit{The table shows the performance metrics for the sensitivity analysis performed on the LSTM model. S\&P 500, FTSE 100, and CAC 40 represent the benchmark Buy\&Hold strategy for each index respectively. S\&P 500 index trading starts on 2005-01-25, FTSE 100 equity index trading starts on 2005-01-13, and CAC 40 equity index trading starts on 2004-12-28. In the base case scenario, the Dropout rate is set to 0.075 and the Batch Size is set to 32. The transaction costs are 0.1\%. The best values are in bold and are bolded with respect to the base case scenario.}}

\linespread{1.5}\selectfont

\hypertarget{lstm-arima-2}{%
\subsection{LSTM-ARIMA}\label{lstm-arima-2}}

Figure 10 and Table 9 present the results of a sensitivity analysis
conducted on the S\&P 500, FTSE 100, and CAC 40 equity indices using the
LSTM-ARIMA model. The results reveal that the highest values of the
modified information ratio (\(IR^{**}\)) are achieved in the base case
scenario for the FTSE 100 and CAC 40 equity indices, as depicted in
Panel E to Panel L.

However, in the case of the S\&P 500 equity index, a higher \(IR^{**}\)
metric is attained for the Long-Only strategy when the dropout rate is
decreased (Panel A) and when the batch size is reduced (Panel B). In
contrast, for the Long-Short strategy, the base case scenario proves to
be the most effective in terms of the \(IR^{**}\) metric.

\pagebreak

\begin{figure}
\centering
\includegraphics{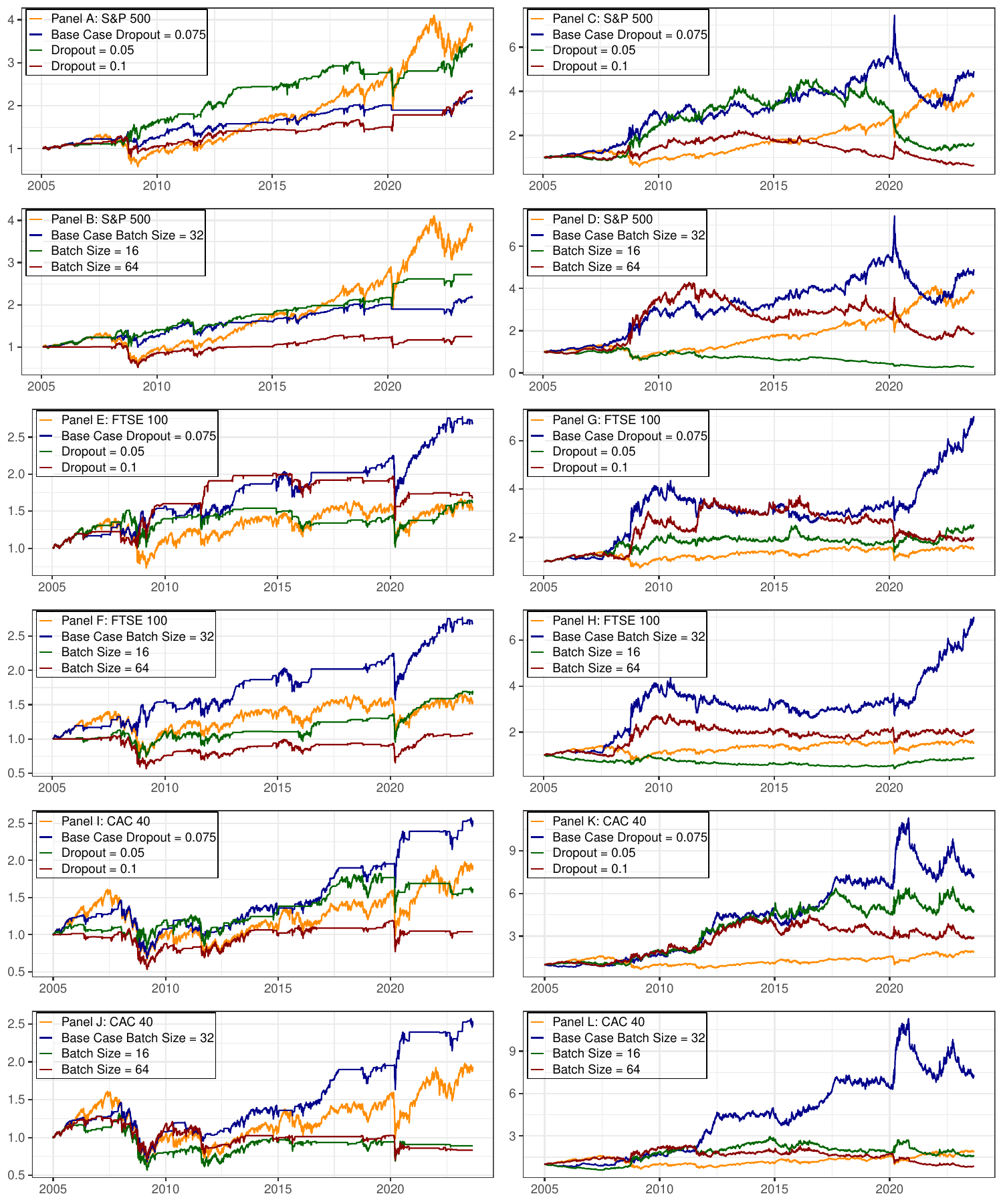}
\caption{LSTM-ARIMA Sensitivity Analysis for S\&P 500}
\end{figure}

\vspace{-0.3cm}

\noindent\linespread{0.65}\selectfont {\scriptsize \textbf{Note}: \textit{The figure presents the equity curves for the sensitivity analysis performed on the LSTM-ARIMA model. S\&P 500, FTSE 100, and CAC 40 represent the benchmark Buy\&Hold strategy for each index respectively. S\&P 500 index trading starts on 2005-01-25, FTSE 100 equity index trading starts on 2005-01-13, and CAC 40 equity index trading starts on 2004-12-28. Each equity curve consists of daily frequency data. The transaction costs are 0.1\%.}}

\linespread{1.5}\selectfont

\pagebreak

\renewcommand{\arraystretch}{0.72}
\fontsize{9}{9.5}\selectfont

\begin{longtable}[]{@{}llcccccc@{}}
\caption{LSTM-ARIMA Sensitivity Analysis performance
metrics}\tabularnewline
\toprule\noalign{}
& & ARC(\%) & ASD(\%) & MD(\%) & MLD & IR*(\%) & IR**(\%) \\
\midrule\noalign{}
\endfirsthead
\toprule\noalign{}
& & ARC(\%) & ASD(\%) & MD(\%) & MLD & IR*(\%) & IR**(\%) \\
\midrule\noalign{}
\endhead
\bottomrule\noalign{}
\endlastfoot
~ & \textbf{S\&P 500} & 7.52 & 19.58 & 56.78 & 1.65 & 38.43 & 5.09 \\
~ & ~ & & & & & & \\
\textbf{Long Only} & Base Case (Dropout = 0.075) & 4.32 & \textbf{11.14}
& 28.95 & \textbf{1.67} & 38.79 & 5.79 \\
\textbf{Panel A: Dropout Rate} & Dropout = 0.05 & \textbf{6.88} & 13.7 &
\textbf{26.63} & 3.92 & \textbf{50.25} & \textbf{12.99} \\
& Dropout = 0.1 & 4.72 & 13.5 & 37.99 & 3.66 & 34.99 & 4.35 \\
& ~ & & & & & & \\
\textbf{Panel B: Batch Size} & Base Case (Batch Size = 32) & 4.32 &
\textbf{11.14} & 28.95 & 1.67 & 38.79 & 5.79 \\
& Batch Size = 16 & \textbf{5.53} & 12.03 & \textbf{27.83} &
\textbf{1.54} & \textbf{45.92} & \textbf{9.11} \\
& Batch Size = 64 & 1.19 & 14.44 & 52.99 & 7.54 & 8.23 & 0.18 \\
& ~ & & & & & & \\
\textbf{Long Short} & Base Case (Dropout = 0.075) & \textbf{8.92} &
\textbf{19.58} & \textbf{56.62} & \textbf{3.44} & \textbf{45.56} &
\textbf{7.18} \\
\textbf{Panel C: Dropout Rate} & Dropout = 0.05 & 2.63 & 19.6 & 71.5 &
6.8 & 13.4 & 0.49 \\
& Dropout = 0.1 & -2.5 & 19.6 & 72.1 & 10.17 & -12.73 & -0.44 \\
& ~ & & & & & & \\
\textbf{Panel D: Batch Size} & Base Case (Batch Size = 32) &
\textbf{8.92} & \textbf{19.58} & \textbf{56.62} & \textbf{3.44} &
\textbf{45.56} & \textbf{7.18} \\
& Batch Size = 16 & -6.58 & \textbf{19.58} & 79.19 & 15.19 & -33.62 &
-2.79 \\
& Batch Size = 64 & 3.35 & 19.59 & 63.83 & 12.32 & 17.12 & 0.9 \\
& ~ & & & & & & \\
~ & \textbf{FTSE 100} & 2.39 & 18.03 & 47.83 & 5.94 & 13.27 & 0.66 \\
~ & ~ & & & & & & \\
\textbf{Long Only} & Base Case (Dropout = 0.075) & \textbf{5.47} & 13.79
& \textbf{30.22} & \textbf{0.91} & \textbf{39.71} & \textbf{7.19} \\
\textbf{Panel E: Dropout Rate} & Dropout = 0.05 & 2.66 & 13.36 & 34.18 &
9.02 & 19.9 & 1.55 \\
& Dropout = 0.1 & 2.85 & \textbf{10.94} & 31.62 & 8.88 & 26.1 & 2.35 \\
& ~ & & & & & & \\
\textbf{Panel F: Batch Size} & Base Case (Batch Size = 32) &
\textbf{5.47} & 13.79 & \textbf{30.22} & 0.91 & \textbf{39.71} &
\textbf{7.19} \\
& Batch Size = 16 & 2.88 & \textbf{13.63} & 36.25 & \textbf{0.73} &
21.15 & 1.68 \\
& Batch Size = 64 & 0.44 & 14.25 & 46.13 & 13.54 & 3.07 & 0.03 \\
& ~ & & & & & & \\
\textbf{Long Short} & Base Case (Dropout = 0.075) & \textbf{10.98} &
\textbf{18.02} & \textbf{40.17} & 10.89 & \textbf{60.92} &
\textbf{16.65} \\
\textbf{Panel G: Dropout Rate} & Dropout = 0.05 & 4.91 & \textbf{18.02}
& 44.39 & 7.59 & 27.24 & 3.01 \\
& Dropout = 0.1 & 3.61 & 18.03 & 51.81 & \textbf{7.56} & 20.03 & 1.4 \\
& ~ & & & & & & \\
\textbf{Panel H: Batch Size} & Base Case (Batch Size = 32) &
\textbf{10.98} & \textbf{18.02} & \textbf{40.17} & \textbf{10.89} &
\textbf{60.92} & \textbf{16.65} \\
& Batch Size = 16 & -0.63 & 18.03 & 60.17 & 14.14 & -3.49 & -0.04 \\
& Batch Size = 64 & 3.98 & 18.04 & 44.77 & 13.2 & 22.08 & 1.96 \\
& ~ & & & & & & \\
~ & \textbf{CAC 40} & 3.52 & 21.44 & 59.16 & 14.04 & 16.43 & 0.98 \\
~ & ~ & & & & & & \\
\textbf{Long Only} & Base Case (Dropout = 0.075) & \textbf{5.02} & 15.43
& 53.65 & 8.33 & \textbf{32.52} & \textbf{3.04} \\
\textbf{Panel I: Dropout Rate} & Dropout = 0.05 & 2.55 & \textbf{15.34}
& \textbf{33.09} & \textbf{4.24} & 16.61 & 1.28 \\
& Dropout = 0.1 & 0.21 & 16.86 & 51.01 & 7.47 & 1.23 & 0.01 \\
& ~ & & & & & & \\
\textbf{Panel J: Batch Size} & Base Case (Batch Size = 32) &
\textbf{5.02} & \textbf{15.43} & 53.65 & \textbf{8.33} & \textbf{32.52}
& \textbf{3.04} \\
& Batch Size = 16 & -0.61 & 17.15 & 56.9 & 15.95 & -3.53 & -0.04 \\
& Batch Size = 64 & -0.95 & 15.78 & \textbf{45.06} & 16.98 & -6.05 &
-0.13 \\
& ~ & & & & & & \\
\textbf{Long Short} & Base Case (Dropout = 0.075) & \textbf{11.06} &
\textbf{21.43} & 39.91 & \textbf{2.91} & \textbf{51.6} &
\textbf{14.29} \\
\textbf{Panel K: Dropout Rate} & Dropout = 0.05 & 8.64 & 21.44 &
\textbf{32.86} & 3.09 & 40.31 & 10.6 \\
& Dropout = 0.1 & 5.7 & 21.44 & 39.43 & 7.27 & 26.57 & 3.84 \\
& ~ & & & & & & \\
\textbf{Panel L: Batch Size} & Base Case (Batch Size = 32) &
\textbf{11.06} & \textbf{21.43} & \textbf{39.91} & \textbf{2.91} &
\textbf{51.6} & \textbf{14.29} \\
& Batch Size = 16 & 2.4 & 21.46 & 48.23 & 9.01 & 11.19 & 0.56 \\
& Batch Size = 64 & -0.89 & 21.44 & 65.76 & 12.72 & -4.15 & -0.06 \\
& ~ & & & & & & \\
\end{longtable}

\normalsize
\vspace{-0.3cm}

\noindent\linespread{0.65}\selectfont {\scriptsize \textbf{Note}: \textit{The table shows the performance metrics for the sensitivity analysis performed on the LSTM-ARIMA model. S\&P 500, FTSE 100, and CAC 40 represent the benchmark Buy\&Hold strategy for each index respectively. S\&P 500 index trading starts on 2005-01-25, FTSE 100 equity index trading starts on 2005-01-13, and CAC 40 equity index trading starts on 2004-12-28. In the base case scenario, the Dropout rate is set to 0.075 and the Batch Size is set to 32. The transaction costs are 0.1\%. The best values are in bold and are bolded with respect to the base case scenario.}}

\linespread{1.5}\selectfont

\pagebreak

\hypertarget{ensembled-ais}{%
\section{Ensembled AIS}\label{ensembled-ais}}

We create an ensemble AIS to diversify the results of our investment
algorithms among all the financial instruments. The idea is that we
invest 1 dollar in each financial instrument and then test the Long-Only
and Long-Short strategy. We assume that the instruments are perfectly
divisible and that we assign a weight of \({\frac{1}{3}}\) to each
equity index. The trading in this case starts on \emph{2005-01-25} and
goes until \emph{2023-08-30}. Figure 11 and Table 10 present the results
for our ensemble AIS. When we aggregate all the equity indices, there is
a notable improvement in our results. We achieve a significantly
enhanced risk-adjusted return (\(IR^{**}\)). The LSTM-ARIMA model
combined with the Long-Short strategy outperforms all other approaches,
yielding an impressive \(IR^{**}\) of \emph{70.54\%}.

\begin{figure}
\centering
\includegraphics[width=1\textwidth]{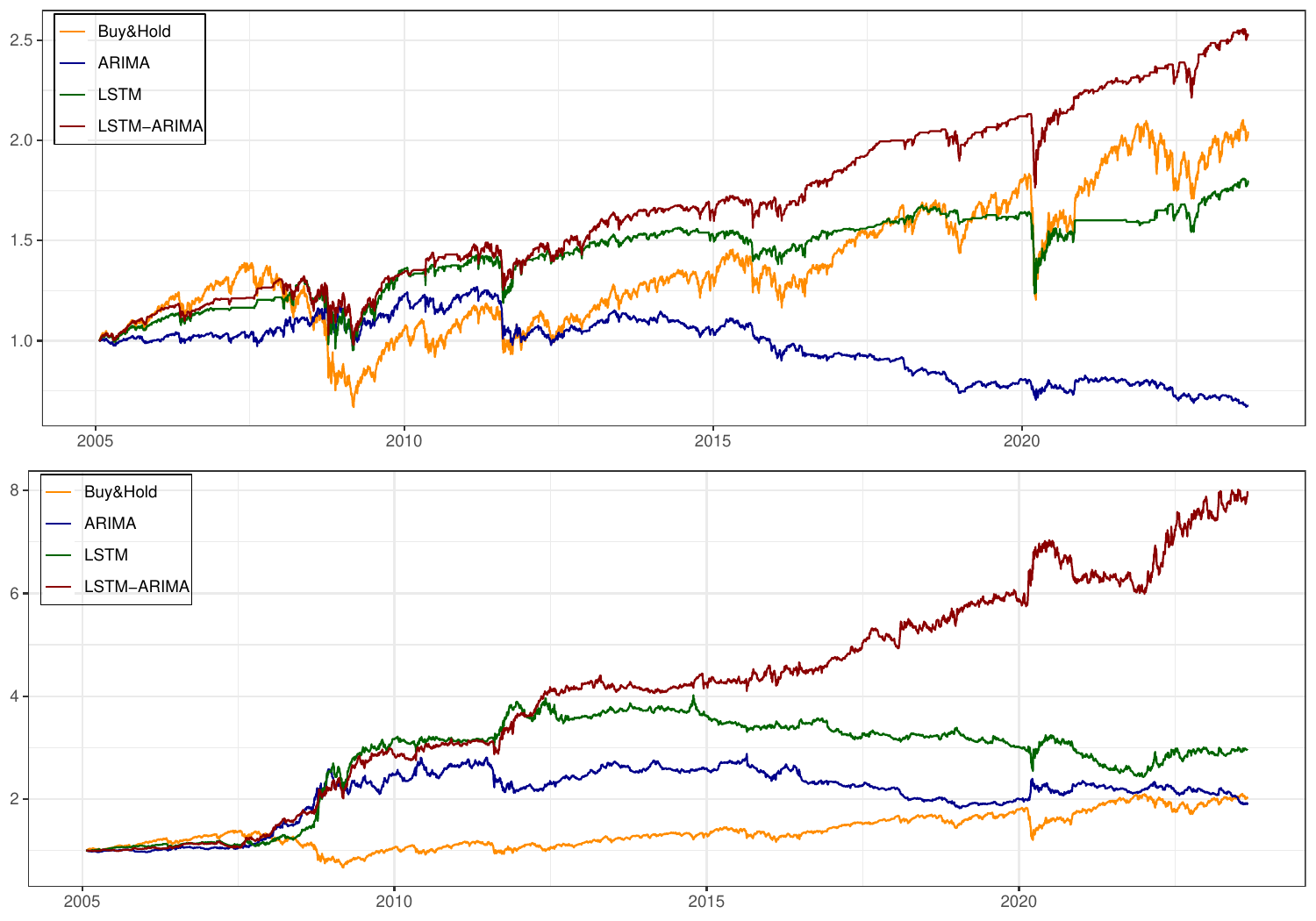}
\caption{The Long-Only and Long-Short Strategy of our ensemble AIS}
\end{figure}

\vspace{-0.4cm}

\noindent\linespread{0.65}\selectfont {\scriptsize \textbf{Note}: \textit{The equity curves are a weighted average of the equity curves of all the equity indices. The weight is equal to 1/3. The first plot presents the equity curve for the Long-Only strategy and the second plot presents the equity curve for the Long-Short strategy. The trading lasts from 2005-01-25 until 2023-08-30. Each equity curve consists of daily frequency data. The transaction costs are 0.1\%.}}

\linespread{1.5}\selectfont

\pagebreak

\renewcommand{\arraystretch}{0.75}
\fontsize{10}{10.5}\selectfont

\begin{longtable}[]{@{}llllllll@{}}
\caption{Performance metrics for ensemble AIS}\tabularnewline
\toprule\noalign{}
& & ARC(\%) & ASD(\%) & MD(\%) & MLD & IR*(\%) & IR**(\%) \\
\midrule\noalign{}
\endfirsthead
\toprule\noalign{}
& & ARC(\%) & ASD(\%) & MD(\%) & MLD & IR*(\%) & IR**(\%) \\
\midrule\noalign{}
\endhead
\bottomrule\noalign{}
\endlastfoot
\textbf{Long Only} & & & & & & & \\
& \textbf{S\&P 500} & 3.92 & 17.43 & 51.87 & 7.7 & 22.48 & 1.7 \\
& ARIMA & -2.09 & 10.93 & 47.24 & 12.47 & -19.09 & -0.84 \\
& LSTM & 3.21 & 11.61 & 27.14 & 4.0 & 27.65 & 3.27 \\
& LSTM-ARIMA & \textbf{5.12} & \textbf{10.43} & \textbf{26.06} &
\textbf{0.42} & \textbf{49.08} & \textbf{9.64} \\
\textbf{Long Short} & & & & & & & \\
& \textbf{S\&P 500} & 3.92 & 17.43 & 51.87 & 7.7 & 22.48 & 1.7 \\
& ARIMA & 3.51 & 12.7 & 36.79 & 8.02 & 27.68 & 2.64 \\
& LSTM & 6.0 & 12.53 & 39.57 & 8.86 & 47.85 & 7.25 \\
& LSTM-ARIMA & \textbf{11.82} & \textbf{11.96} & \textbf{16.57} &
\textbf{1.87} & \textbf{98.86} & \textbf{70.54} \\
\end{longtable}

\normalsize
\vspace{-0.2cm}

\noindent\linespread{0.65}\selectfont {\scriptsize \textbf{Note}: \textit{The trading lasts from 2005-01-25 until 2023-08-30. The transaction costs are 0.1\%. The best strategy is the one that holds the highest Modified Information Ratio ($IR^{**}$). Columns with the best corresponding values are denoted in bold.}}

\linespread{1.5}\selectfont

\hypertarget{conclusion}{%
\section{Conclusion}\label{conclusion}}

This study aimed to create a strategy using LSTM-ARIMA that performs
better than the individual algorithms. To assess the performance of
tested approaches, we created three algorithmic investment strategies
based on ARIMA, LSTM, and LSTM-ARIMA models. We conducted hyperparameter
tuning using a random search. The walk-forward optimization was applied
to perform the model training and backtesting. The best model was chosen
based on the conditions presented in section 4.7. Next, we generated
buy/sell signals using the condition explained in Section 4.8. The
algorithmic investment strategy was tested on three different equity
indices: S\&P 500, FTSE 100, and CAC 40 on daily frequency data between
the period of January 2000 and August 2023. The algorithm predicted the
next day's closing price based on the historical data and was classified
as a regression problem.

According to our initial hypotheses, the LSTM-ARIMA model was expected
to outperform other algorithms in the majority of cases. The LSTM-ARIMA
model indeed outperformed all the other algorithms in all the cases, the
summary can be read in section 5.2 where it is outlined which strategy
performed the best based on the \(IR^{**}\) metric. Therefore, we
conclude that we have no grounds to reject our hypotheses. The answers
to the research questions stated in the \(Introduction\) are presented
below:

\pagebreak

\begin{quote}
\textbf{RQ1. Are the investment algorithms robust to changes in the
asset?} \newline Our hybrid model outperforms all the other models
across all equity indices, though with varying performance metrics. The
varying performance metrics across different assets indicate that our
algorithms are not robust to changes in the asset. This suggests that
the algorithms' performance is sensitive to the specific characteristics
and dynamics of each asset, requiring further modifications or
adaptations to improve their effectiveness across different assets.
\newline  \textbf{RQ2. Does LSTM-ARIMA perform better than the models
individually?} \newline Based on the findings presented in Section 5.1
and the performance metrics provided in Tables 2, 3, and 4, it can be
concluded that LSTM-ARIMA outperforms the other models examined.
\newline  \textbf{RQ3. Are the algorithmic investment strategies robust
to changes in the model hyperparameters?} \newline In section 6, it
becomes apparent that modifications to the hyperparameters have an
impact on the results, indicating a lack of robustness. This implies
that the model's performance is sensitive to the specific choices made
for hyperparameter settings. \newline \textbf{RQ4. Does the Long-Only or
Long-Short strategy outperform the Buy\&Hold?} \newline Based on the
analysis presented in Section 5.1 and the data provided in Tables 2, 3,
and 4, several findings can be observed. In the case of the S\&P 500
equity index, it has been observed that both the Long-Only and
Long-Short strategies implemented with the LSTM-ARIMA model, along with
the Long-Short strategy implemented with the ARIMA model outperform the
Buy\&Hold strategy. In the case of the FTSE 100 equity index, the
Long-Only and Long-Short strategies implemented with the LSTM-ARIMA and
LSTM model outperform the Buy\&Hold strategy. In the case of the CAC 40
equity index, it has been noticed that both the Long-Only and Long-Short
strategies implemented with the LSTM-ARIMA model, along with the
Long-Only strategy implemented with the LSTM model outperform the
Buy\&Hold strategy.
\end{quote}

This study has made a valuable contribution to the existing literature
by introducing a hybrid approach that combines modern forecasting
models, such as LSTM and ARIMA, for algorithmic investment strategies.
Previous research has explored this hybrid approach in various domains.
For instance, Bali et al. (2020) utilized LSTM-ARIMA to forecast wind
speed, Fan et al. (2021) optimized the hybrid model for well production
forecasting, Dave et al. (2021) employed it to forecast exports in
Indonesia, and Arnob et al. (2019) used the hybrid approach for
forecasting the Dhaka stock exchange (DSE). However, our study stands
out by applying this model specifically to algorithmic investment
strategies, a relatively uncommon application for this hybrid approach.

In the base case scenario outlined in Section 5, our hybrid approach
outperformed all the models, aligning with our expectations. However,
through the course of our sensitivity analysis, it became apparent that
there is room for further enhancements by adjusting the dropout rate and
the batch size.

There are several potential directions to expand upon this research.
Firstly, it would be valuable to investigate whether utilizing returns
as inputs instead of the closing price impacts the outcomes. Secondly,
the sensitivity analysis revealed noteworthy improvements in the results
by reducing the dropout rate and the batch size S\&P 500 equity index
Long-Only strategy. Therefore, conducting a more comprehensive
sensitivity analysis by examining a broader range of dropout rates and
sizes would be beneficial. Thirdly, it is worth evaluating the changes
in results when considering a binary cross-entropy problem. Finally, it
is important to explore the use of a specific threshold that indicates
when to change the position in both the Long-Only and Long-Short
strategies.

\fontsize{8}{8.5}\selectfont

\hypertarget{refs}{}

\end{document}